\documentclass[11pt]{article}

\def\gtwid{\mathrel{\raise.3ex\hbox{$>$\kern-.75em\lower1ex\hbox{$\sim
$}}}}
\def\vio{\mathrel{\hbox{$E$\kern-.60em\hbox{$/
$}}}}
\newcommand{\mhsig}{\ensuremath{m_{H_{\rm sig}}}}
\newcommand{\rhsiggg}{\ensuremath{R_{H_{\rm sig}}^{\gamma\gamma}}}
\newcommand{\rhsigzz}{\ensuremath{R_{H_{\rm sig}}^{ZZ}}}
\newcommand{\rhigg}{\ensuremath{R_{H_i}^{\gamma\gamma}}}

\newcommand{\rhtwogg}{\ensuremath{R_{H_2}^{\gamma\gamma}}}

\newcommand{\phlam}{\ensuremath{\phi_\lambda}}
\newcommand{\phkap}{\ensuremath{\phi_\kappa}}

\newcommand{\phtri}{\ensuremath{\phi_{A_0}}}

\usepackage{graphicx}
\usepackage{amsmath}
\usepackage{subfig}
\usepackage{float}

\begin{document}

\thispagestyle{empty}
\begin{flushright}
SHEP-13-07\\
April 2013
\end{flushright}
\vspace*{2.0cm}
\begin{center}
{\Large \bf 125 GeV Higgs boson signal within the Complex NMSSM}\\ 
\vspace{.3in}
{\large S. Moretti$^{a,\dagger}$, S. Munir$^{b,\dagger}$ and
  P. Poulose$^{c,\dagger}$ } \\[0.25cm]
{\sl $^{a}$ School of Physics \& Astronomy, \\
University of Southampton, Southampton SO17 1BJ, UK.} \\[0.25cm]
{\sl $^b$ National Centre for Nuclear Research,
  Ho{\. z}a 69, 00-681 Warsaw, Poland. }\\[0.25cm]
{\sl $^c$ Department of Physics, \\ 
IIT Guwahati, Assam 781039, India. } \\[0.25cm]
\end{center}
\vspace{0.3in}
\begin{abstract}
\noindent
While the properties of the 125 GeV Higgs boson-like particle observed by
the ATLAS and CMS collaborations 
are largely compatible with those predicted for
the Standard Model state, significant
deviations are present in some cases. We, therefore, test the
viability of a Beyond the Standard Model scenario based on Supersymmetry, 
the CP-violating Next-to-Minimal Supersymmetric Standard Model, 
against the corresponding experimental observations.
Namely, we identify possible model configurations in which one of its Higgs bosons
is consistent with the LHC observation and evaluate the role of the
explicit complex phases in both the mass and
di-photon decay of such a Higgs boson. Through a detailed
analysis of some benchmark points corresponding to each of these configurations, we highlight the
impact of the CP-violating phases on the model predictions compared to the CP-conserving case.
\end{abstract}
\vskip 50mm
\noindent {\footnotesize $^\dagger$E-mails:\\
{\tt 
{S.Moretti@soton.ac.uk},\\
{SMunir@fuw.edu.pl},\\
{Poulose@iitg.ernet.in}.
}
}

\newpage
\section{Introduction}\label{intro}

In July 2012, the CMS and ATLAS experimental collaborations at the
Large Hadron Collider (LHC) announced the observation of a
new boson \cite{:2012gk,:2012gu}, consistent with a Higgs particle, the last undiscovered
object in the Standard Model (SM). The initial results were based on
data corresponding to integrated luminosities of $5.1$\,fb$^{-1}$ taken at $\sqrt{s} = 7$\,TeV and 5.3\,fb$^{-1}$ at $8$\,TeV and the search was performed in six decay modes: $H \to \gamma
\gamma$, $ZZ$, $Z\gamma$, $WW$, $\tau^+ \tau^-$ and $b\bar{b}$. A $\sim$5$\sigma$ excess of
events with respect to the background was clearly observed in the
first and second of these decay modes, while the remaining ones
yielded exclusion limits well above the SM expectation. Both collaborations have
since been regularly updating their
findings \cite{LHCresults1cms,LHCresults1atlas, LHCresults2cms,LHCresults2atlas,LHCresults3}, improving the mass and 
(so-called) `signal strength' measurements. 

In these searches, the magnitude of a possible signal is characterized by the production cross section
times the relevant Branching Ratios (BRs) relative to the SM
expectations in a given Higgs boson decay channel $X$, denoted by $R(X)= \sigma/\sigma_{\rm SM}
\times{\rm BR}(X)/{\rm BR}_{\rm SM}(X)$ (i.e., the signal strength).
According to the latest results released by the two collaborations
after the collection of $\sim$20 fb$^{-1}$ of data 
\cite{LHCresults2cms,LHCresults2atlas,LHCresults3}, 
a broad resonance compatible with a 125\,GeV
signal is now also visible in the $WW \rightarrow 2l2\nu$ decay channel. The mass of the
observed particle is still centered around 125 GeV but the measured
values of its signal strength in different channels have changed considerably compared to the
earlier results. These values now read \\
\hspace*{1cm}$R(\gamma \gamma) = 0.78 \pm 0.28$,~~~
$R(ZZ) = 0.91^{+0.3}_{-0.24}$, ~~~
$R(WW) = 0.76\pm0.21$ \\
\noindent at CMS, and\\ 
\hspace*{1cm}$R(\gamma \gamma) = 1.65\pm0.35$, ~~~
$R(ZZ) = 1.7\pm0.5$, ~~~
$R(WW) = 1.01\pm 0.31$\\
\noindent at  ATLAS. The bulk of the event rates comes from the gluon-gluon fusion channel \cite{WJSZK}.
Furthermore, the signal has also been corroborated by Tevatron analyses \cite{HiggsEvidenceTevatron},
covering the $b\bar b$ decay mode only, with the Higgs boson stemming from associated
production with a $W$ boson \cite{WJSZK}. However, there the
comparisons against the SM Higgs boson rates are biased by much larger experimental errors. 

If the current properties of the observed particle are confirmed 
after an analysis of the full 7 and 8\,TeV data samples from the LHC, they
will not only be a clear signature of a Higgs boson, but also a
significant hint for possible physics beyond the SM.
In fact, quite apart from noting that the current data are not entirely 
compatible with SM Higgs boson production rates, 
while the most significant LHC measurements point to a mass 
for the new resonance around 125\,GeV
the Tevatron excess in the $b\bar b$ channel points to a range
between 115\,GeV and 135\,GeV. While the possibility that the SM
Higgs boson state has any of such masses would be merely a coincidence (as its mass is a free parameter),
in generic Supersymmetry (SUSY) models the mass of the lightest Higgs boson with SM-like
behavior is naturally confined to be less than 180\,GeV or so
\cite{CPNSH}. The reason is that SUSY, in essence, relates trilinear Higgs boson and gauge couplings, so that the former are of the same size as the latter, in turn implying such a small
Higgs boson mass value. Therefore, the new LHC results could well be perceived as being in favor of some low energy SUSY realisation.

Several representations of the latter have recently been studied in
connection with the aforementioned LHC and Tevatron data,
 including the Minimal Supersymmetric Standard Model
 (MSSM) \cite{mssm125} (also the constrained 
version \cite{cmssm125} of it, in fact), the
Next-to-Minimal Supersymmetric Standard Model
(NMSSM) \cite{Ellwanger:2011aa,nmssm-set1,Kowalska:2012gs}, 
the E$_6$-inspired Supersymmetric Standard
Model (E$_6$SSM) \cite{e6ssm125} and the (B-L) Supersymmetric Standard
Model ((B-L)SSM) \cite{blssm125}.  All of these scenarios can yield
a SM-like Higgs boson with mass around 125\,GeV and most of them can
additionally explain the excesses in the signal strength measurements in the di-photon channel. 

Another approach to adopt in order to test the viability of SUSY solutions to the LHC Higgs boson data is 
to consider the possibility of having CP-violating (CPV) phases (for a
general review of CP Violation, see Ref. \cite{Ibrahim:2007fb}) in (some
of) the SUSY parameters. These phases can substantially modify Higgs boson phenomenology 
in both the mass spectrum and production/decay rates at the LHC
\cite{cpv1,cpv2,cpvall,cpv0}, while at the same time providing a solution
to electroweak baryogenesis \cite{Cirigliano:2009yd}. In the context of the LHC, the impact
of CPV phases was emphasized 
long ago in Ref. \cite{past-papers1,past-papers2} and revisited recently in Ref. \cite{Chakraborty:2013si} following the Higgs boson discovery.
In all such papers though, CPV effects were studied in the case of the MSSM.

In this paper, we consider the case of similar CPV effects in the NMSSM.
In particular, we study the possibility to have Higgs boson signals with mass around 125\,GeV in the
CPV NMSSM, which are in agreement with the aforementioned LHC 
data as well as the direct search constraints on sparticle masses from
LEP and LHC. We also investigate the dependence of the feasible 
CPV NMSSM signals on the mass of  the Higgs boson as well as its couplings to both 
the relevant particle and sparticle states entering the model 
spectrum, chiefly, through the decay of the former into a $\gamma\gamma$ pair. 
We thus aim at a general understanding of how such observables are
affected by the possible complex phases explicitly entering the Higgs sector of the
next-to-minimal SUSY Lagrangian.

The paper is organized as follows. In the next section, we will 
briefly review the possible explicit CPV phases in the Higgs sector of the
NMSSM. In Sec. \ref{sec:params} we will outline the independent CPV NMSSM
parameters and the methodology adopted to confine our attention to the
subset of them that can impinge on the LHC Higgs boson data. 
In the same section, we further investigate the possible numerical values
of the complex parameters after performing scans of the low energy CPV NMSSM
observables compatible with the LEP and LHC constraints on Higgs boson and
SUSY masses. 
In Sec. \ref{sec:results} we present our results on the Higgs boson 
mass spectrum as well as signal rates in connection 
with the LHC. Finally, we conclude in Sec. \ref{sec:summa}.

\section{CPV phases in the Higgs sector of the NMSSM}
\label{sec:CPV}

The CPV phases appearing in the Higgs potential of the
NMSSM at tree-level \cite{nmssm125b} can be divided into three categories:
\begin{enumerate}
\item $\theta$ and $\varphi$: the spontaneous CPV phases
of the vacuum expectation values (vevs) of the up-type Higgs doublet $H_u$ and the Higgs singlet
$S$, respectively, with respect to the down-type Higgs doublet $H_d$; 
\item $\phi_\lambda$ and $\phi_\kappa$: the phases of the Higgs boson
trilinear couplings $\lambda$ and $\kappa$; 
\item $\phi_{A_\lambda}$ and $\phi_{A_\kappa}$: the phases of the
trilinear soft terms $A_\lambda$ and $A_\kappa$. 
\end{enumerate}

As explained in \cite{Cheung:2010ba,Graf:2012hh}
the phases in category 3.  above
are determined by the minimisation conditions of the Higgs potential
with respect to the three Higgs fields. Furthermore, assuming
vanishing spontaneous CPV phases in category 1. (and real SM Yukawa couplings), the only actual physical phases
appearing in the tree-level Higgs potential are those in category 2 as the difference
$\phi_\lambda - \phi_\kappa$.
Beyond the Born approximation, the phases of the trilinear couplings
$A_t$, $A_b$, and $A_\tau$ also enter the Higgs sector through
radiative corrections from the third generation squarks and stau
(assuming negligible corrections from the first two
generations). Also, in the one-loop effective potential, \phlam\ 
can contribute independently from \phkap. The 
complete one-loop Higgs mass matrix can be found in Refs. \cite{Cheung:2010ba,Graf:2012hh,Funakubo:2004ka}.
Here we only reproduce the tree-level Higgs as well as sfermion mass
matrices in Appendix A.

The $5\times5$ Higgs mass matrix ${\cal M}^2_H$, defined in the basis ${\bf H}^T\equiv\left(
H_{dR},\,H_{uR},\,S_R,\,H_I,\,S_I\right)$ (after $\beta$-rotating the
$6\times 6$ matrix to isolate the Goldstone mode), is diagonalized with a unitary matrix $O$ to
yield five mass eigenstates as
\begin{equation}
\label{eq:h-states}
\left(H_{dR},\,H_{uR},\,S_R,\,H_I,\,S_I \right)^T \ = \
O~\left( H_1,\,H_2,\,H_3,\,H_4,\,H_5\right)^T,
\end{equation}
where $O^T{\cal M}_H^2 O = {\sf diag}(m^2_{H_1},\,m^2_{H_2},\,m^2_{H_3},\,m^2_{H_4},\,m^2_{H_5})$
 in order of increasing mass. For a nonzero value of any of the
phases listed above, these mass eigenstates become CP indefinite due
to scalar-pseudoscalar mixing. Moreover, these CPV phases not only
affect the masses of the Higgs states but also their decay widths,
since the Higgs boson couplings to various particles  
are proportional to the elements of the
unitary matrix $O$ (see, e.g., Refs. \cite{CPSuperH,Cheung:2011wn}). Additionally, alterations in the masses of light 
neutralinos and charginos, in particular, due to the phases in category
1 above, can also have an indirect impact on the BRs of the Higgs bosons into SM particles. 

The decay widths and BRs of the Higgs boson in the NMSSM with
CPV phases can be calculated using the methodology implemented in Ref.
\cite{CPSuperH}. Explicit expressions for Higgs boson couplings and widths 
in the CPV NMSSM can be found in Ref. \cite{Munir:2013dya}, which follows the
notation of \cite{nmssm125a}. These widths and BRs can
then be used to obtain the signal strength of the $\gamma\gamma$
channel (also called {\it reduced} di-photon cross section), \rhigg,
defined, for a given Higgs boson, $H_i$, as 
\begin{eqnarray}
\label{eq:Rxsct1}
R_{H_i}^{\gamma\gamma} =  \frac{\sigma(gg\rightarrow
  H_i)}{\sigma(gg\rightarrow H_{\rm SM})}\times \frac{{\rm BR}(H_i
  \rightarrow \gamma\gamma)}{{\rm BR}(H_{\rm SM} \rightarrow \gamma\gamma)},
\end{eqnarray}
where $H_{\rm SM}$ implies a SM Higgs boson with
the same mass as $H_i$. In terms of the {\it reduced} couplings, $C_i(X)$ (couplings of $H_i$ with
respect to those of $H_{\rm SM}$), Eq.\,(\ref{eq:Rxsct1}) can be approximated by
\begin{eqnarray}
\label{eq:Rxsct2}
R_{H_i}^{\gamma\gamma} = [C_i(gg)]^2[C_i(\gamma\gamma)]^2\sum_X
\frac{\Gamma^{\textrm{total}}_{h_\textrm{SM}}}{\Gamma^{\textrm{total}}_{H_i}}
\end{eqnarray}
where $\Gamma^{\textrm{total}}_{h_\textrm{SM}}$ denotes the total
width of $H_{\rm SM}$. 

\section{Model parameters and methodology}
\label{sec:params}

In light of the recent LHC discovery of a SM Higgs boson-like particle we
scan the parameter space of the CPV Higgs sector
of the NMSSM using a newly developed {\sc fortran} code. 
In our scans the LEP constraints on the model Higgs bosons are imposed in 
a modified fashion; i.e., they have to be satisfied by the scalar and
pseudoscalar components of all the CP-mixed Higgs bosons. Also imposed are
 the constraints from the direct searches of the third-generation
 squarks, stau and the light chargino at the LEP. We point out here that,
 in the CP-conserving (CPC) limit, the Higgs boson mass and BRs have been compared
 with those given by NMSSMTools \cite{Tools} and have been found
 to differ from the latter by $\sim$1\% and $\sim$5\% at the most, respectively.
Although no limits from $b$-physics or from relic density measurements 
have been imposed we confine ourselves to the regions
the parameter space regions which have been found to comply with 
such constraints (see, e.g., Ref. \cite{Kowalska:2012gs}). 

We study the effects of the CPV phases described in the previous section on
the mass and di-photon signal rate of a Higgs boson predicted by the
model that is compatible with the Higgs boson discovery data from the LHC. 
In particular, we consider the three most likely scenarios specific to the CPV
NMSSM that comply with the latter. In our analysis, we assume minimal Supergravity
(mSUGRA)-like unification of the soft parameters at the SUSY-breaking energy
scale, such that 
\begin{center}
 $M_0 \equiv M_{Q_3} = M_{U_3} = M_{D_3} = M_{L_3} = M_{E_3} = M_{\rm SUSY}$, \\
 $M_{1/2} \equiv 2M_1 = M_2 =  \frac{1}{3} M_3$,  \\
 $A_0 \equiv A_t = A_b = A_{\tau}$, \\
\end{center}
where $M^2_{\widetilde{Q}_3},\,M^2_{\widetilde{U}_3},\,M^2_{\widetilde{D}_3}$
and $M^2_{\widetilde{L}_3},\,M^2_{\widetilde{E}_3}$ are the soft
SUSY-breaking squared masses
of the third-generation squarks and sleptons, respectively. 
These parameters are then fixed to their optimal values based on
earlier studies \cite{Ellwanger:2011aa,Kowalska:2012gs} in order to 
minimize the set of scanned parameters. We
then focus only on the effects of the Higgs sector parameters, which
include the dimensionless Higgs boson couplings $\lambda$ and $\kappa$ along with
their phases \phlam\ and \phkap, as well as the soft
SUSY-breaking 
parameters $A_\lambda$ and $A_\kappa$. From outside the Higgs sector, we
only analyze the effect of the variation of the unified CPV phase of
the third-generation trilinear couplings, $\phi_{A_0}$ ($\equiv \phi_{A_t}=\phi_{A_b}=\phi_{A_\tau}$).   

Before we discuss the three scenarios mentioned above, we note
that the two heaviest Higgs boson mass
eigenstates $H_4$ and $H_5$ always correspond to the interaction eigenstates $H_{uR}$
and $H_I$ in Eq.\,(\ref{eq:h-states}).\footnote{Implying that after diagonalization
of the Higgs mass matrix, e.g., $H_4$ sits in the position
corresponding to $H_{uR}$ in Eq.\,(\ref{eq:h-states}) before ordering by mass, even
though evidently it contains components of the other Higgs fields
also. In particular, a SM-like $H_i$ contains adequate components of
both $H_{uR}$ and $H_{dR}$ to have SM-like couplings to fermions and
gauge bosons.}
Hence, a scenario is defined by the Higgs state that
conforms to the LHC observations, out of 
the three light mass eigenstates, $H_1$, $H_2$ and $H_3$, and by the correspondence between the
latter and the interaction eigenstates $H_{dR}$, $S_R$ and $S_I$. 
However, note that such a definition is adopted only so that a distinction between different scenarios can be made conveniently.
Evidently, the behavior of the `observed' Higgs boson, $H_{\rm sig}$, with the CPV phases in a given scenario is a combined result of the set of parameters yielding that scenario rather than of its position among the
mass-ordered Higgs states. 
The criteria for choosing the ranges of the scanned model parameters 
as well as the values of the non-Higgs-sector SUSY parameters thus depend on the 
scenario under consideration and are explained in the following.\\

\noindent \underline{Scenario\,1:} In this scenario the lightest Higgs state, $H_1$, is
 the SM-like one and corresponds to $H_{dR}$, while $H_2$ and $H_3$
 correspond to $S_R$ and $S_I$, respectively. The 
requirement of obtaining a down-type Higgs state with mass close to 125 GeV and with SM-like 
couplings necessitates large soft SUSY masses and $A_0$. The values of $\mu_{\rm eff}$ ($\equiv \lambda
s$, where $s$ is the vev of $S$) and the gaugino masses
are found to be in best agreement with the relic density constraints \cite{Kowalska:2012gs},
giving a neutralino with a large Higgsino
component as the lightest SUSY particle. Further, $\lambda$ and
$\kappa$ are chosen such that there is enough 
mixing of the doublet with the singlet Higgs boson so as to allow an
$H_1$ with the correct mass while keeping its couplings close to their
SM values. We test two cases for this scenario, corresponding to two 
representative values of the parameter $\tan\beta$ ($\equiv v_u/v_d$,
where $v_u$ and $v_d$ are the vevs of $H_u$ and $H_d$, respectively), which is fixed to
8 in Case\,1 and to 15 in Case\,2. \\

\noindent \underline{Scenario\,2:} This scenario is defined by
 the SM-like $\sim$125\,GeV Higgs boson being the second lightest Higgs boson,
$H_2$, of the model. There are two possibility entailing
 such a scenario. It can be $H_{d_R}$-like with a large singlet component,
 in which case it has \rhigg\ SM-like or bigger,
 as shown in \cite{Ellwanger:2011aa}. We refer to this
 possibility as Case\,1 of this scenario. It requires relatively large values of $\lambda$ and
$\kappa$, small values of the parameters $A_\lambda$ and $A_\kappa$
and moderate values of soft SUSY-breaking parameters. 
For Case\,2 of this scenario, we take a slightly different region of
the parameter space which yields a
$H_2$ that is again $H_{dR}$-like but with a much smaller singlet
component, so that it has \rhtwogg\ around the SM expectation.
Therefore, heavy unified soft squark mass and/or trilinear coupling 
are required in this Case, but a light soft gaugino mass is
preferred. $\lambda$ can be small to intermediate while
$\kappa$ is always small. Finally, in this scenario $H_1$ and $H_3$ are $S_R$- and $S_I$-like, respectively. \\

\noindent \underline{Scenario\,3:} There also exists the possibility
that the observed $\sim$125\,GeV Higgs boson
is the $H_3$ of the model which corresponds to $H_{dR}$, while both 
$S_R$- and $S_I$-like Higgs bosons are lighter. Such a scenario
can be realized for very fine-tuned
ranges of the parameters $A_\lambda$ and $A_\kappa$ for a given
$\tan\beta$ value, with large soft squark
and gaugino mass parameters preferred. Note that in this Case
 the $S_I$-like $H_3$ of Case\,2 of scenario\,2 turns into $H_2$ by
 becoming lighter than the $H_{dR}$-like state which, consequently, turns into
$H_3$. These two cases thus overlap slightly in terms of the
relevant parameter space of the model. \\

Note here that we do not consider a scenario with the
$\sim$125\,GeV Higgs boson corresponding to the $S_I$ interaction eigenstate,
since the pure (or nearly pure) pseudoscalar hypothesis is disfavored by the CMS
Higgs boson analyses
\cite{Chatrchyan:2012jja,Djouadi:2013qya}. Moreover, in scenarios 2
and 3 above the masses of $H_2$ and $H_3$ can lie very close to
each other. In fact, these two Higgs bosons can be almost
degenerate in mass near 125\,GeV, in which case the signal observed at the LHC
should be interpreted as a superposition of individual peaks due to each of
them. However, in the NMSSM, particularly in the presence of CPV
phases, more than one possibilies with mass degenerate Higgs bosons may
arise (see, e.g., \cite{Gunion:degen} and \cite{Munir:2013wka}).
Such possibilities warrant a dedicated study of their own, which is
currently underway, and in this
article we have, therefore, not taken any of them into account. For
correctness of our results, we have thus imposed the condition
of nondegeneracy during our scans, so that only those points are 
passed for which no other Higgs boson apart from the signal Higgs boson under 
consideration lies inside the mass range of interest, defined in
the next section. Values of the fixed parameters as well as ranges of the variable
parameters for all the above scenarios are given in Table\,\ref{tab:range_par}.  

\begin{table}[t]
\begin{center}
\begin{tabular}{|l|c|c|c|c|c|}
\hline
Scenario & 1,\,Case\,1 & 1,\,Case\,2  & 2,\,Case\,1 & 2,\,Case\,2 & 3\\
\hline
\multicolumn{6}{|l|}{Fixed parameters} \\
\hline
$M_0$\,(TeV) & \multicolumn{2}{|c|}{5} & 0.8 & 3 & 3 \\
$M_{1/2}$\,(TeV)  & \multicolumn{2}{|c|}{3} & 0.35 & 0.35 & 1.5 \\
$- A_0$\,(TeV)  & \multicolumn{2}{|c|}{10} & 1 & 4 & 4 \\
$\mu_{\rm eff}$\,(TeV) &  \multicolumn{2}{|c|}{1} & 0.14 & 0.14 & 0.14 \\
$\tan\beta$ & 8 & 15 & 1.9 & 20 & 10 \\
\hline
\multicolumn{6}{|l|}{Scanned parameters} \\
\hline
$\lambda$  & \multicolumn{2}{|c|}{0.01 -- 0.1} & 0.5 -- 0.6 & 0.01 --
0.3 & 0.1 -- 0.3 \\
$\kappa$  & \multicolumn{2}{|c|}{0.1 -- 0.3} & 0.3 -- 0.4 & 0.01 --
0.1 & 0.05 -- 0.1 \\
$A_\lambda$\,(TeV) & \multicolumn{2}{|c|}{1.5 -- 3} & 0.14 -- 0.2 & 0.2
-- 0.6 & 0.95 -- 1.05 \\
$- A_\kappa$\,(TeV) & \multicolumn{2}{|c|}{1 -- 4} & 0.2 -- 0.25 & 0.1
-- 0.3 & 0.07 -- 0.09 \\
\hline
\end{tabular}
\caption{Input parameters of the CPV NMSSM and their numerical values adopted in our analysis.}
\label{tab:range_par}
\end{center}
\end{table}

\section{Scans and results}
\label{sec:results}

We perform scans for each of the scenarios described earlier requiring
the mass of $H_{\rm sig}$ (i.e., of $H_1$ in Scenario\,1, $H_2$ in Scenario\,2 and $H_3$ in Scenario\,3) to lie in the range
124\,GeV\,$<m_{H_{\rm sig}} <$\,127\,GeV.\footnote{We thus use the central mass
  measurement of 125.5 GeV in accordance with the CMS results.} 
We additionally impose the condition
$\rhsiggg>0.5$ on the signal Higgs boson.
Furthermore, to each Case in a given Scenario corresponds a set of three scans,
such that in each of the scans only one of the following CPV phases is varied: \\

\noindent (i) $\phi_{\kappa}$, ~~~(ii) $\phi_{\lambda}$, ~~~(iii) $\phi_{A_0}$, \\

\noindent while fixing the others to $0^\circ$. Each scan thus
checks the effect of a different CPV source at the tree-level and/or 
beyond. In each scan we vary the relevant phase 
in steps of 1$^\circ$ between $0^\circ$ and  $180^\circ$.




The measurements of the Electric Dipole Moment (EDM) of the electron, neutron, and various atoms 
\cite{Baker:2006ts,Commins:2007zz,Griffith:2009zz} put constraints on the allowed values of \phlam\ and \phtri. However, the trilinear couplings of squarks and sleptons contribute to the EDMs only at the two-loop level, and their phases are thus rather weakly constrained. One can, furthermore, assign very heavy soft masses to the sfermions of first two generations in order to  minimize the effect of \phtri\ on the EDMs, as pointed out in earlier studies for the MSSM \cite{EDM-MSSM}. In fact, such
constraints can be neglected altogether by arguing that the phase
combinations occurring in the EDMs can be different from the ones
inducing Higgs boson mixing \cite{EDM_const}. 
 The phase of $\kappa$, in contrast, has been found to be virtually
unconstrained by the EDM measurements \cite{Cheung:2011wn,Graf:2012hh}. 

Below we present our results separately for each of the five cases investigated.
For evaluating the effect of the phases on \mhsig\ and \rhsiggg\
qualitatively, we choose a set of four representative points (RPs), referred
to as RP1, RP2, RP3, and RP4 in the following, for every Case. As
explained in Sect.\,\ref{sec:CPV}, at the tree-level the only independent CPV
phase entering the Higgs mass matrix is the difference
$\phlam-\phkap$. At one-loop level, although \phlam\ can appear
separately from \phkap, the contribution from the corresponding terms is
much smaller than the tree-level dependence on
$\phlam-\phkap$. Furthermore, since only $\cos(\phlam-\phkap)$ appears in the
diagonal CP-even and CP-odd blocks and only $\sin(\phlam-\phkap)$ in the
CP-mixing block (see Appendix A) of the Higgs mass matrix, the
mass eigenstates show a very identical behavior when
either of these two phases is varied while fixing the other to 0$^\circ$. Only small
differences arise for very large values of \phlam\ and \phkap\ due to the higher-order
corrections. Therefore, our RP1 for a given Case
corresponds to a point for which the effect of \phlam\ (and
equivalently \phkap) on \mhsig\ is maximized for that
Case. Similarly, RP3 is chosen such that the variation in \mhsig\ is
maximal with \phtri, since this phase only appears at the
one-loop level and can potentially cause a behavior different from
that due to the tree-level CPV phase. The dependence on \phtri\ can, however,
be expected to show an identical behavior across all cases, 
as it is largely independent of other Higgs sector parameters (except
$\tan\beta$), which is indeed what we will observe in our results below.  

As already noted, the CPV phases also affect
the Higgs boson decay widths into fermions and gauge bosons, through
the elements of the Higgs mixing matrix $O$. On the other hand, in the decays of a Higgs
boson into two lighter Higgs bosons, the tree-level phase,
$\phlam-\phkap$, enters directly while the phase \phtri\ also enters
through the one-loop CP-odd tadpole conditions at
one-loop \cite{Cheung:2010ba,Munir:2013dya}. RP2 and RP4 are, therefore, points with the
largest effect on \rhsiggg\ due to the variation in \phlam/\phkap\ and
\phtri, respectively, observed in our scans.
Note that in the discussion below the description of
the behavior of a given RP may not be equally applicable to
all other good points, since it is chosen only so as to understand the
maximum possible impact of a given phase and to highlight some
potentially distinguishing features of different cases and Scenarios.

\subsection{Scenario\,1:}

\begin{figure}[p]
\centering
\subfloat[]{%
\includegraphics*[angle=-90, scale=0.4]{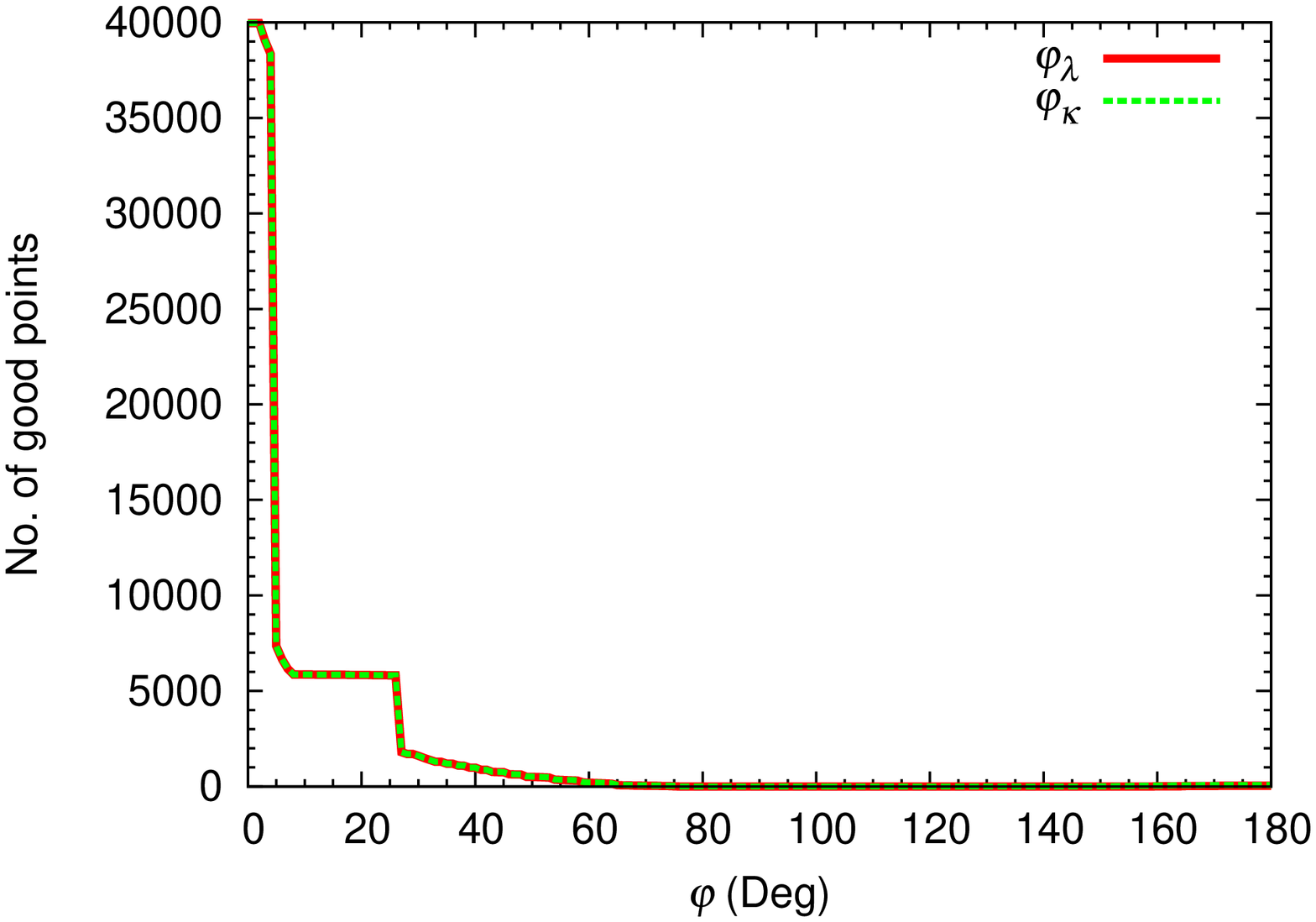}
}
\subfloat[]{%
\includegraphics*[angle=-90, scale=0.4]{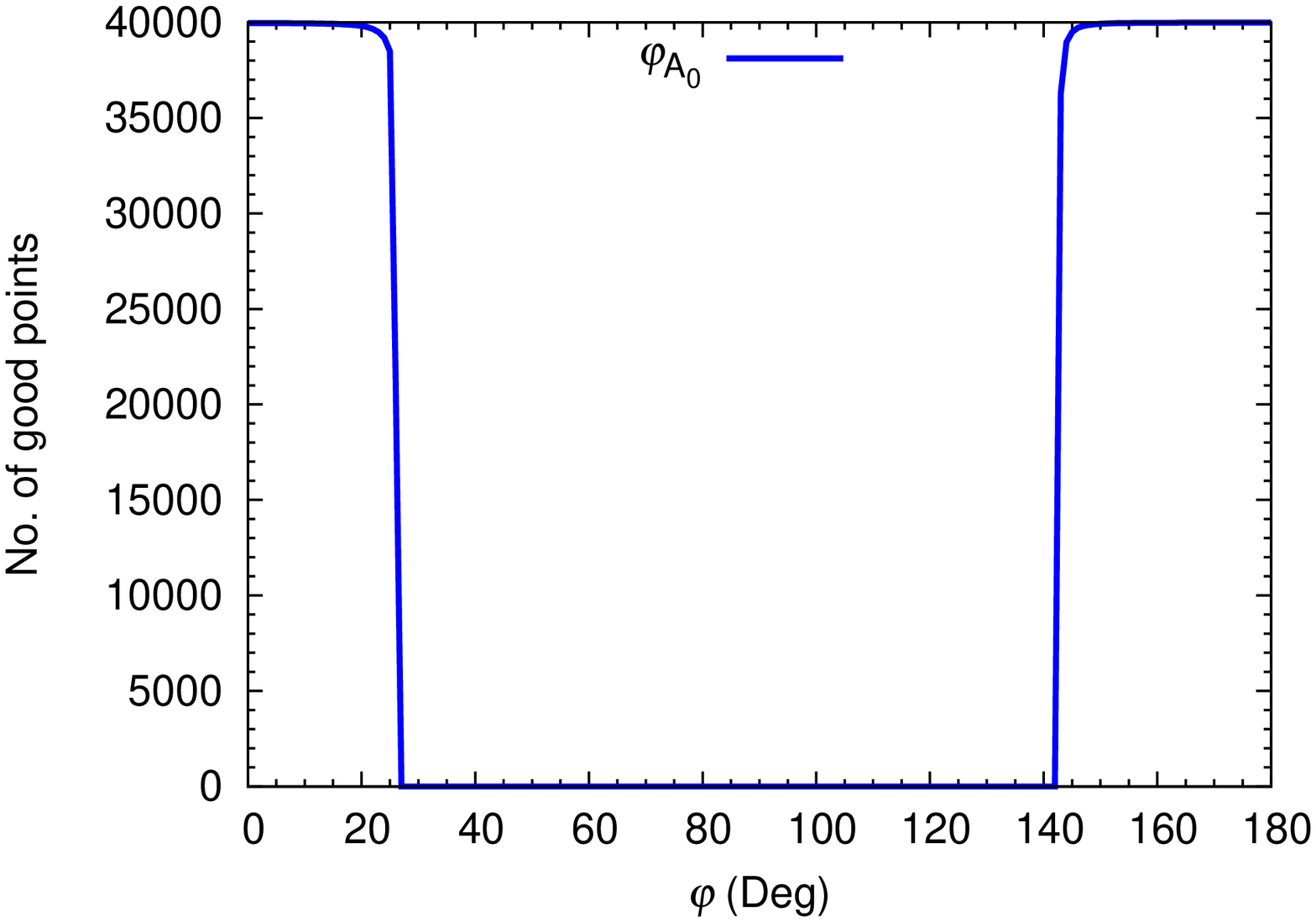}
}

\centering
\subfloat[]{%
\includegraphics*[angle=-90, scale=0.4]{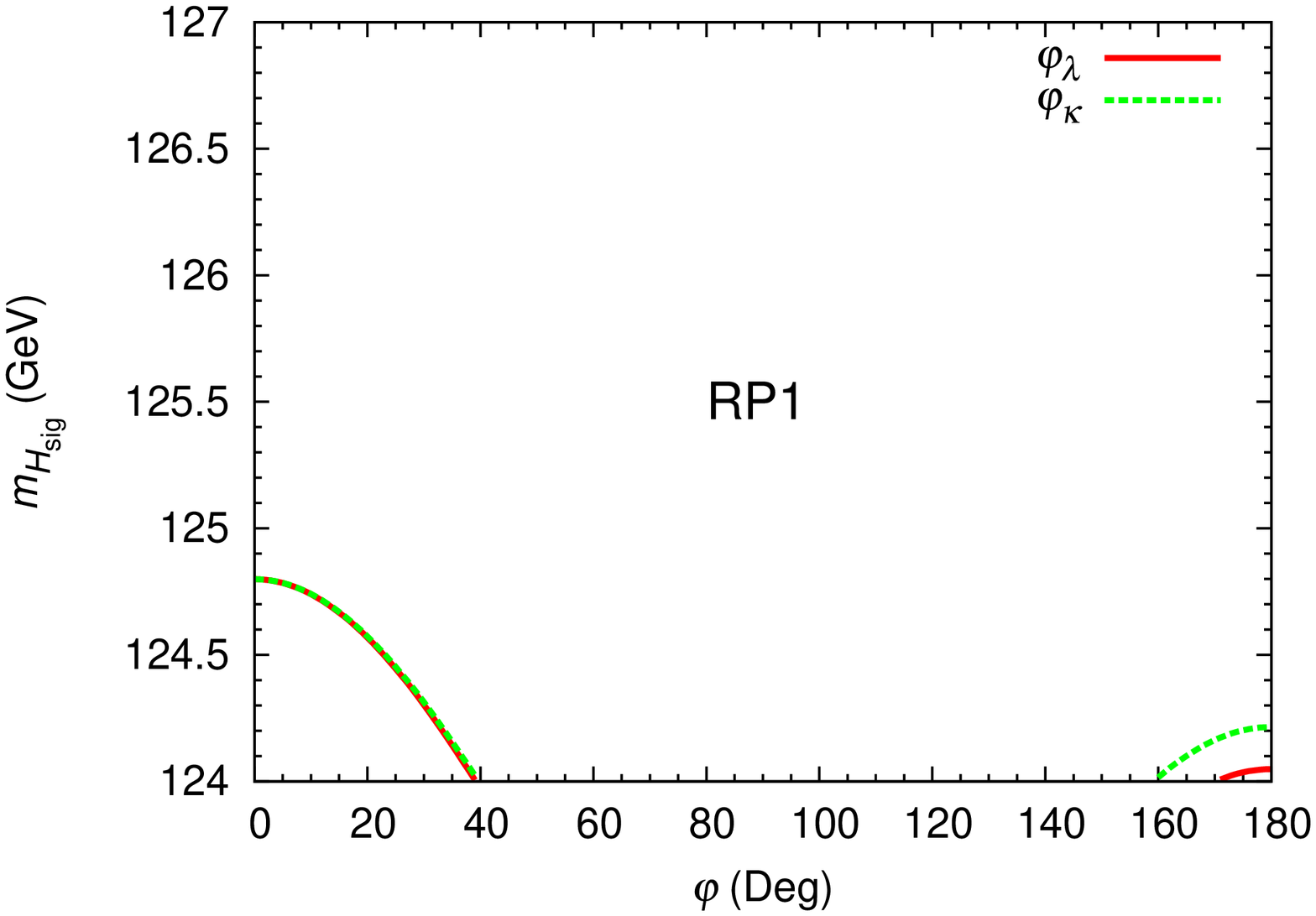}
}
\subfloat[]{%
\includegraphics*[angle=-90, scale=0.4]{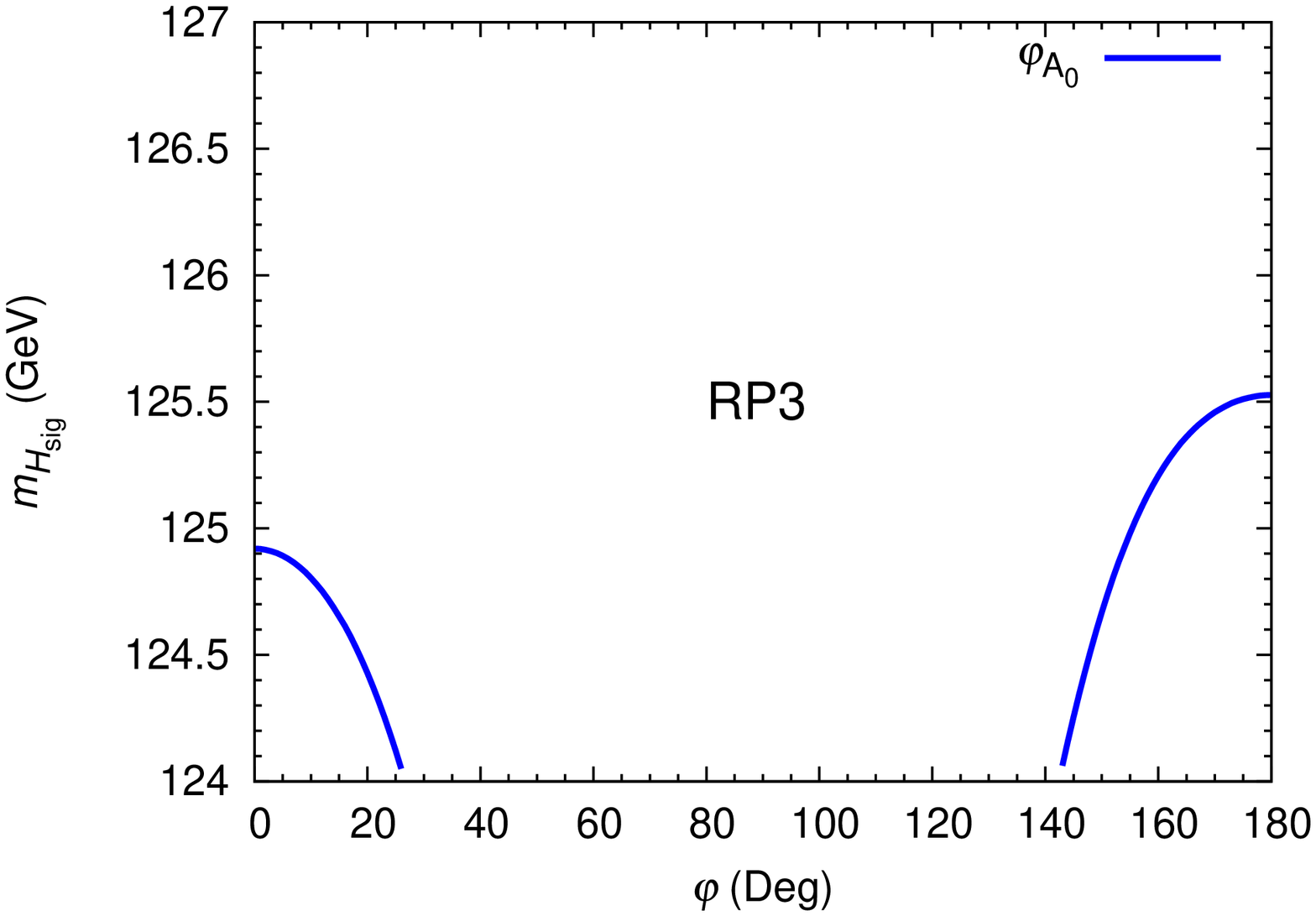}
}

\centering
\subfloat[]{%
\includegraphics*[angle=-90, scale=0.4]{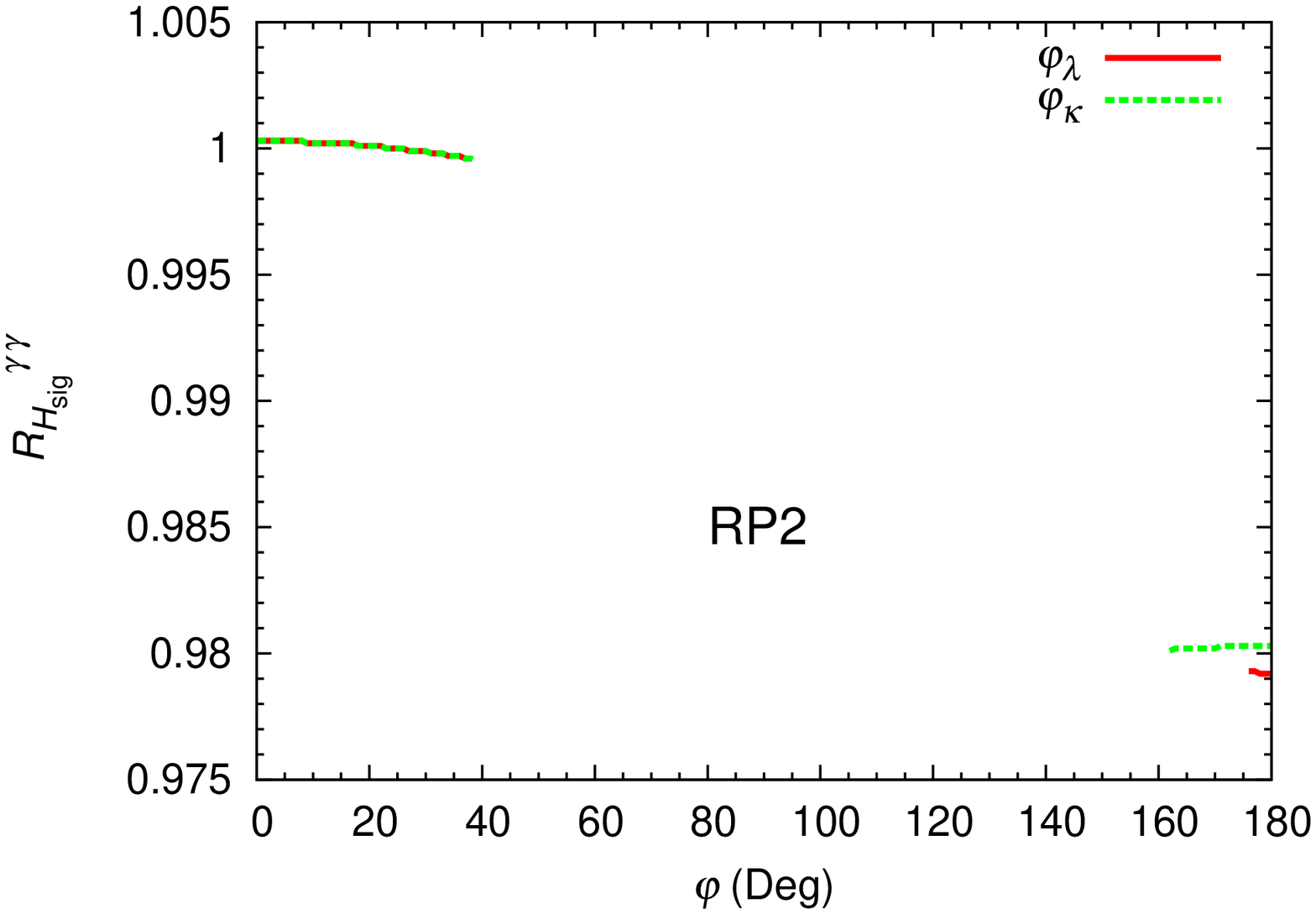}
}
\subfloat[]{%
\includegraphics*[angle=-90, scale=0.4]{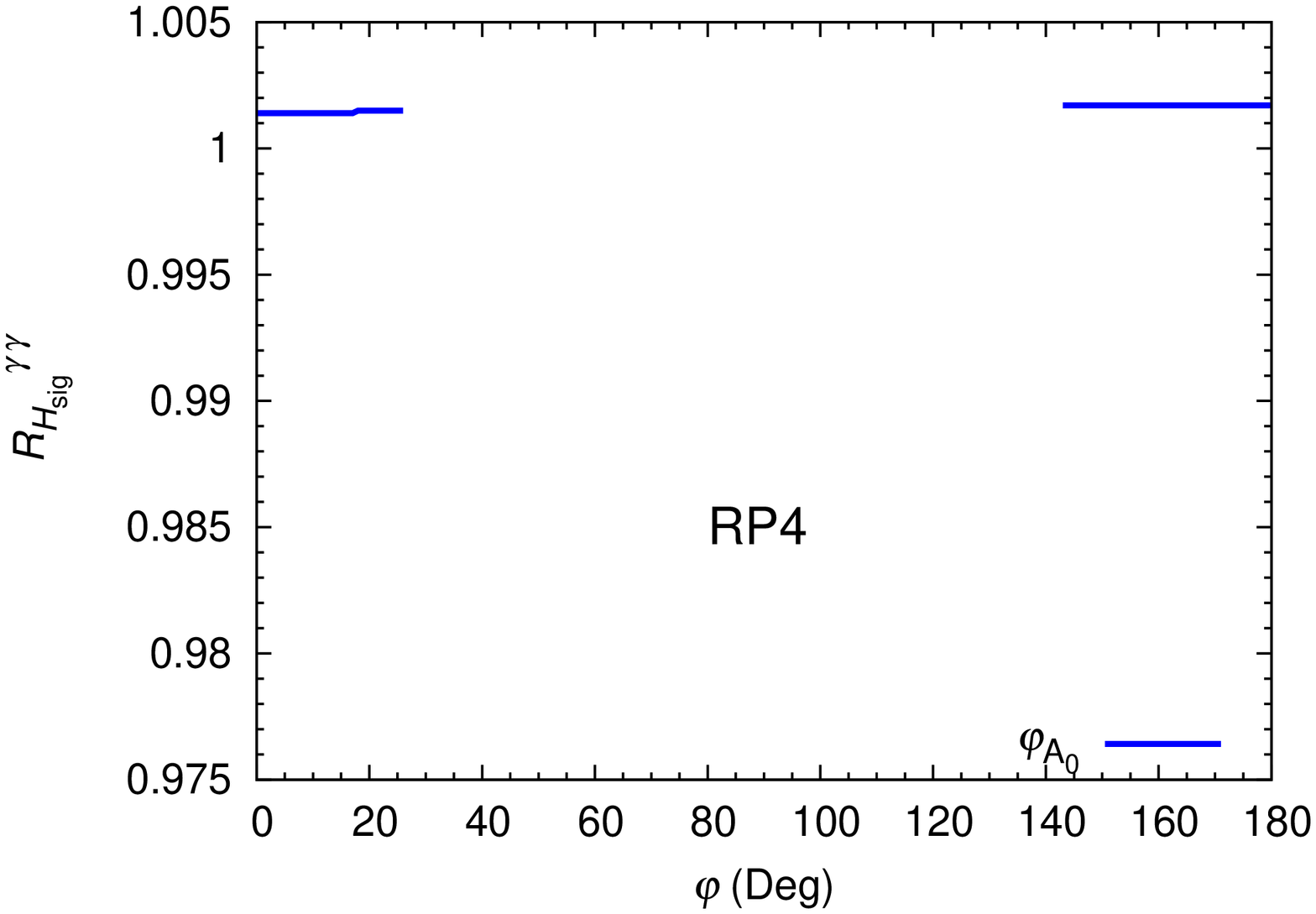}
}

\caption{Distributions of good points, $m_{H_{\rm sig}}$ and $R^{\gamma\gamma}_{H_{\rm sig}}$ for Scenario\,1, Case\,1.}
\label{fig:S1C1}
\end{figure}

\underline{Case\,1:} In Fig.\,\ref{fig:S1C1}a we show, for the small
$\tan\beta$ case of this Scenario, the variation in the
number of good points, i.e., points surviving the conditions imposed
on \mhsig\ and \rhsiggg, with
varying \phlam\ and \phkap. The number
of surviving points first falls slowly with increasing 
\phlam/\phkap\ and then abruptly for $\phlam/\phkap = 5^\circ$ after
which it remains almost constant for a while before falling further.
The number of surviving points reduces to 0 for \phlam\ and
\phkap\ larger than 75$^\circ$ and 77$^\circ$, respectively. However,
 very few, $\sim$10,
surviving points reemerge for \phlam\ and \phkap\ larger than
155$^\circ$ and 150$^\circ$, respectively, although it is not apparent
from the figure. The drop in the number of points is not continuous since there are other parameters, $\lambda$, $\kappa$,
$A_\lambda$ and $A_\kappa$, which are also scanned over for every
value of a given phase. Moreover, the number of good points 
evidently depends on the conditions on \mhsig\ and
\rhsiggg\ so that while both of these may be satisfied for one value
of a phase one of these may be violated for the next. Although, as we
shall see below, \mhsig\ is almost always mainly responsible for the
drop in the number of good points. 

Note that the CPC case is also subject to the conditions on \mhsig\ and
\rhsiggg\ and, on account
of being defined relative to this Case, the number
of good points does not represent all possible solutions for all values of the
phases. Thus, it is likely that the CPC case for a given parameter
set falls outside the defined ranges of \mhsig\ and/or
 \rhsiggg, but the conditions on these are satisfied for a different value of a particular
phase. Such a value of the phase can thus result in a considerable
number of good points which would be absent in the CPC case. Nevertheless,
the aim here is to give an estimate of the effect of the
CPV phases on the number of good points relative to the CPC case,
rather than to present a truly holistic picture. Fig.\,\ref{fig:S1C1}b shows
the variation in the number of surviving points with that in \phtri. Contrary
to the case of \phlam/\phkap, for this phase the number of surviving
points falls abruptly to 0 when \phtri\ crosses $25^\circ$ and then
rises again when \phtri\ reaches $142^\circ$. The values of other parameters corresponding to each 
RP for this Case are given below.
\begin{table}[h]
\begin{center}
\begin{tabular}{c|c|c|c|c}
 Point & $\lambda$ & $\kappa$ & $A_\lambda$\,(GeV) & $A_\kappa$\,(GeV) \\
\hline
RP1 & 0.091 & 0.13 & 1833 & -1000 \\
\hline
RP2 & 0.091 & 0.13 & 1667 & -1000 \\
\hline
RP3 & 0.1 & 0.26 & 3000 & -4000 \\
\hline
RP4 & 0.01 & 0.23 & 2667 & -1000\\
\end{tabular}
\end{center}
\end{table}

Fig.\,\ref{fig:S1C1}c for RP1 verifies our statement above that the condition
on \mhsig\ is the one mainly responsible for the drop in the number of
good points. This is particularly true for this Case due to the fact that
maximum $m_{H_1}$ obtainable for
$\phi_\kappa/\phi_\lambda=0^\circ$ already lies not far above the
allowed lower limit and the former falls further 
with increasing values of these phases. Note in the figure that for
very large values of \phlam\ and \phkap\ $m_{H_1}$ rises above the
lower limit again, more so for the latter than the former. Also, as
\phlam\ and \phkap\ reach very large values,
the lines corresponding to these two phases start deviating slightly from
each other, which is caused by different higher order contributions from
either of these phases, as noted earlier.
The variation in $m_{H_1}$ with increasing $\phi_{A_0}$ is relatively
sharp, as seen in Fig.\,\ref{fig:S1C1}d for RP3. 
This is due to the fact that in order to reach values up to 125\,GeV,
the mass of $H_1$ strongly relies on the trilinear coupling $A_0$ and
is consequently also more sensitive to its phase. Also, while
\mhsig\ falls initially with increasing \phtri, it appears to reach a
minimum for a certain (intermediate) value of \phtri\ after which it starts
rising again. This rise is in fact faster than the earlier drop and as a result $m_{H_1}$ for RP2 is larger for
$\phtri=180^\circ$ than for $\phtri=0^\circ$.

Fig.\,\ref{fig:S1C1}e corresponds to RP2 and
 shows the dependence of \rhsiggg\ on
\phlam\ and \phkap\ for this Case. We
note that \rhsiggg\ falls very slowly with an increase in the value of either of
these phases, deviating from the CPC case only at a percent level for
$\phlam/\phkap \sim 180^\circ$ but still staying very
SM-like. The large break in the line corresponds to those values of
the phases for which $m_{H_1}$ for this RP falls below the allowed
range in analogy with RP1 in Fig.\,\ref{fig:S1C1}c above. The observed
behavior of \rhsiggg\ is due to the fact that with increasing
values of \phlam/\phkap\ while 
BR$(H_1\rightarrow b\bar{b})$ drops and BR$(H_1\rightarrow
\gamma\gamma)$ rises,
 there is a drop in $\Gamma(H_1\rightarrow gg)$ also
(see Eq.\,(\ref{eq:Rxsct1})), resulting in an overall (slight)
reduction in
\rhsiggg. Conversely, for very large values of \phlam\ and \phkap\ the
drop in \rhsiggg\ is even more significant because BR$(H_1\rightarrow
b\bar{b})$ (BR$(H_1\rightarrow\gamma\gamma)$) is slightly larger
(smaller) than its value in the CPC case. Finally, Fig.\,\ref{fig:S1C1}f for
RP4 shows that \rhsiggg\ 
has negligible dependence on \phtri, since for the range of the latter
allowed by the condition on \mhsig\ the total decay width of $H_1$ as
well as BR$(H_1\rightarrow b\bar{b})$ fall negligibly. \\

\begin{figure}[p]
\centering
\subfloat[]{%
\includegraphics*[angle=-90, scale=0.4]{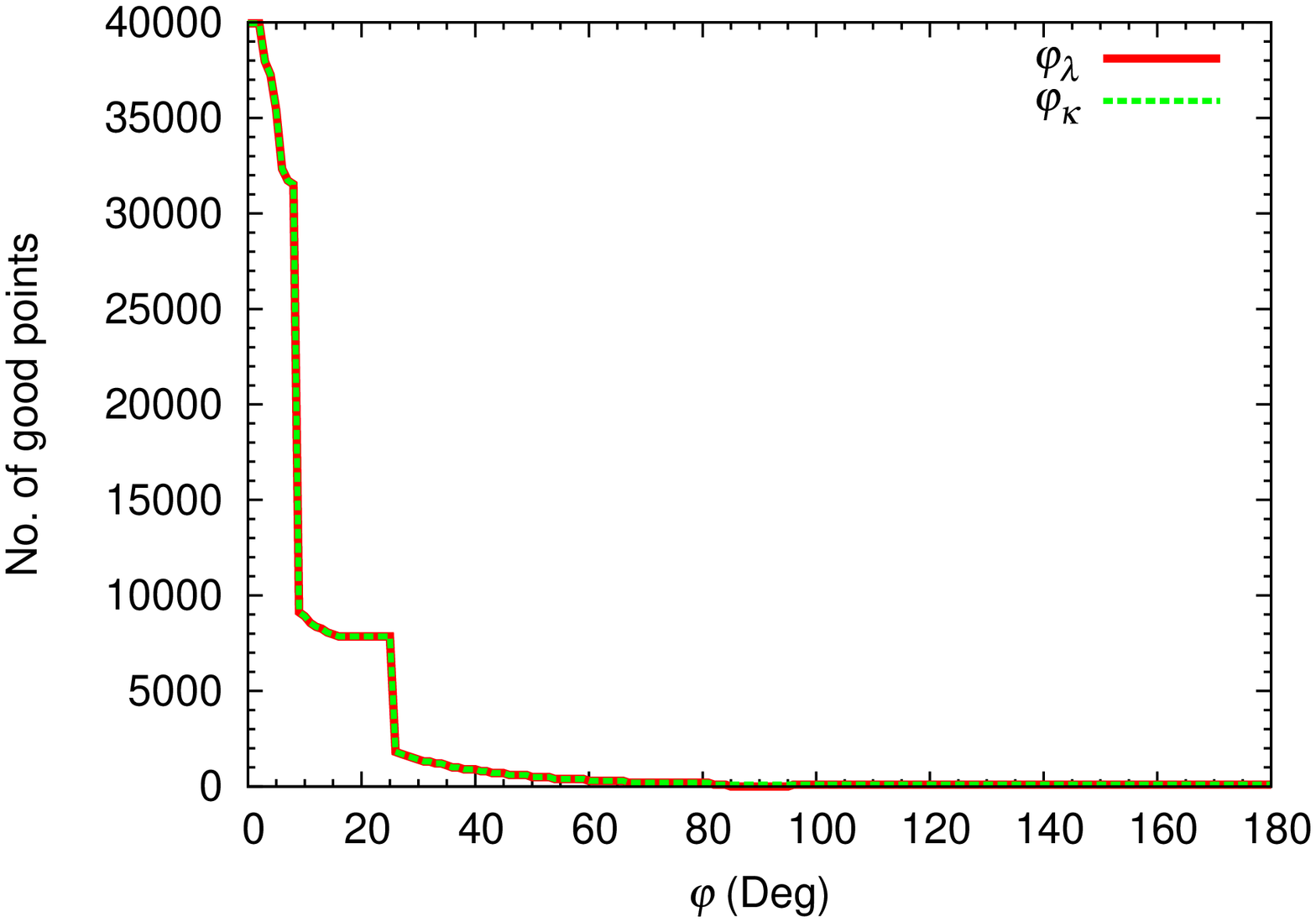}
}
\subfloat[]{%
\includegraphics*[angle=-90,scale=0.4]{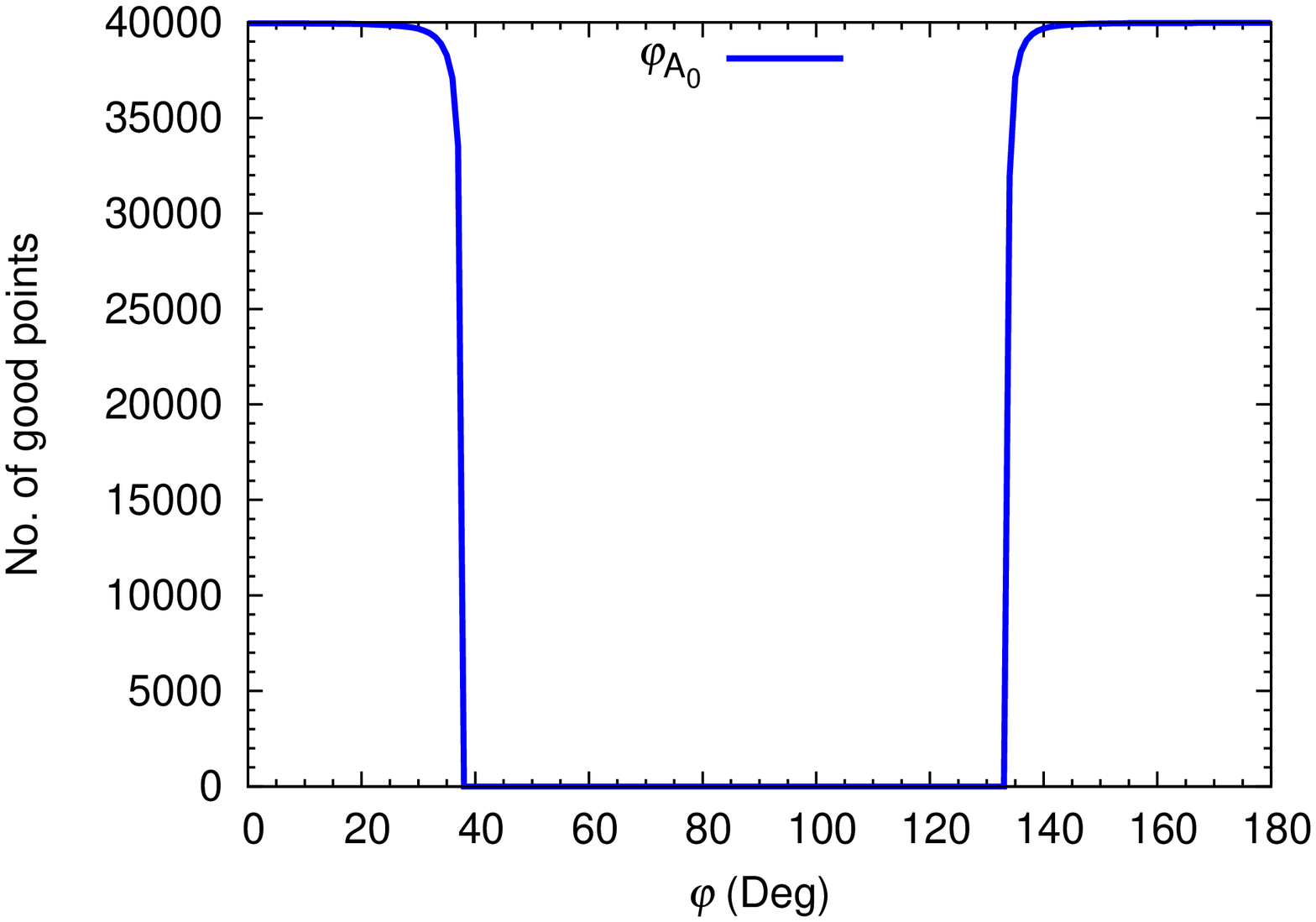}
}

\centering
\subfloat[]{%
\includegraphics*[angle=-90, scale=0.4]{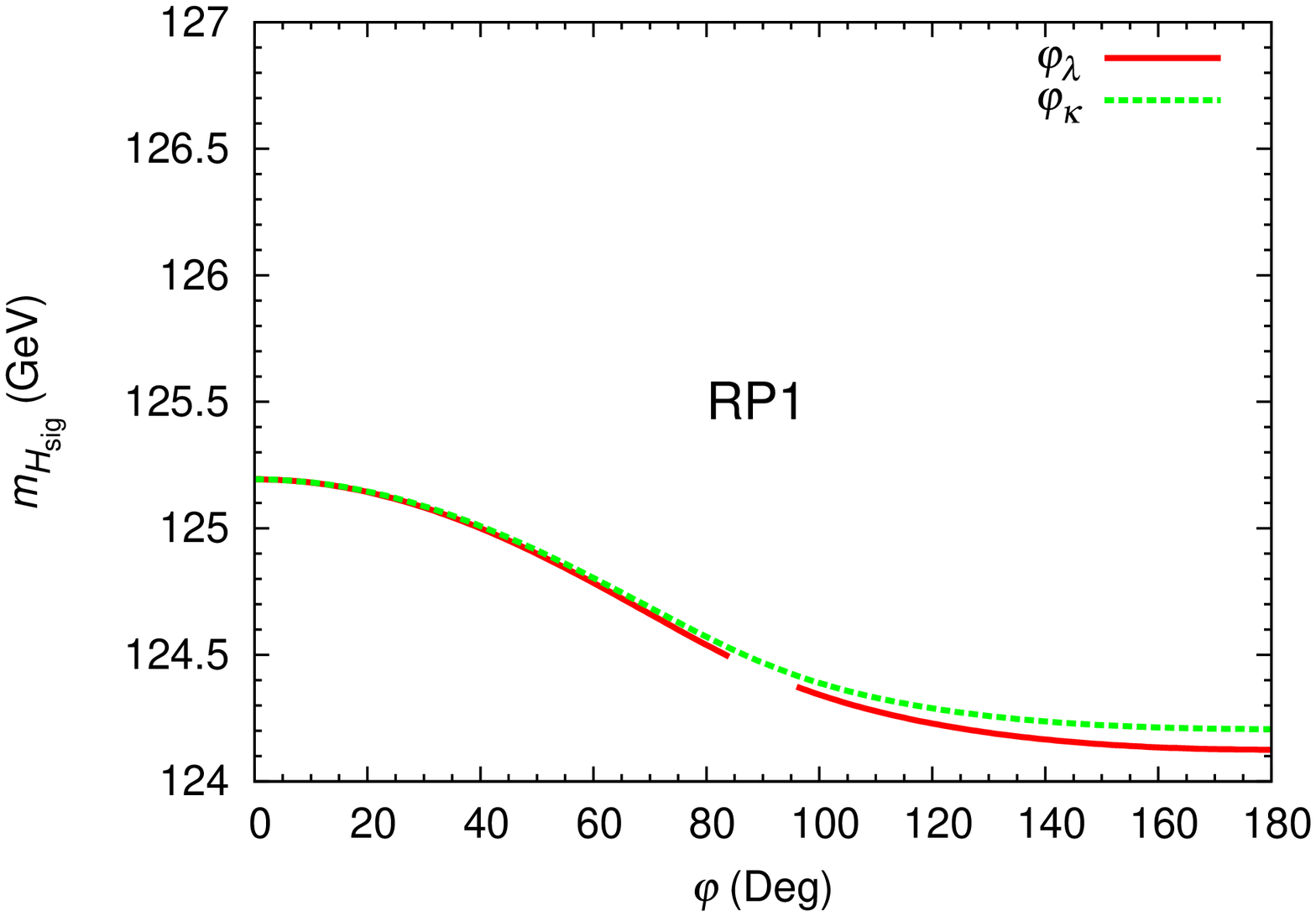}
}
\subfloat[]{%
\includegraphics*[angle=-90, scale=0.4]{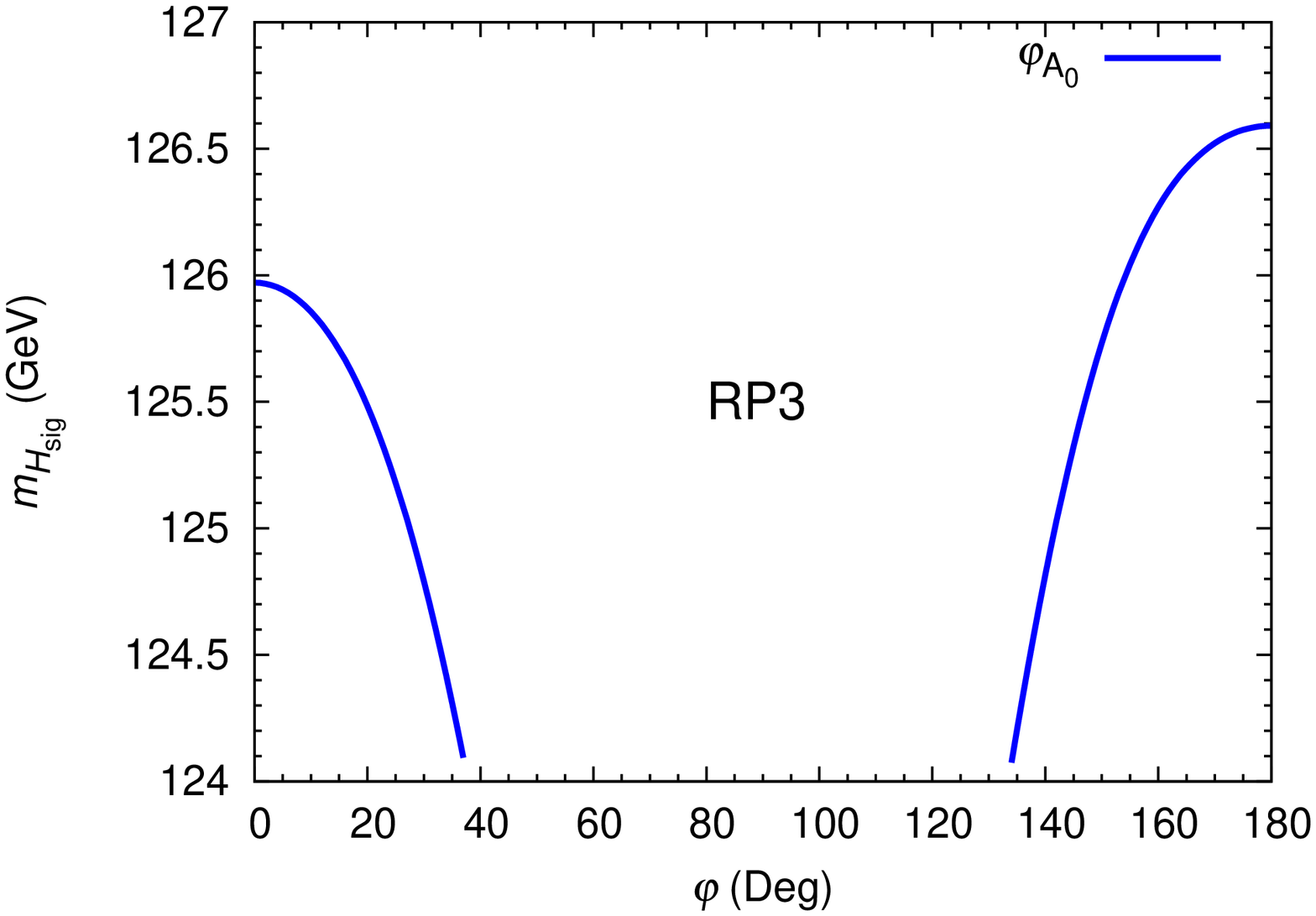}
}

\centering
\subfloat[]{%
\includegraphics*[angle=-90, scale=0.4]{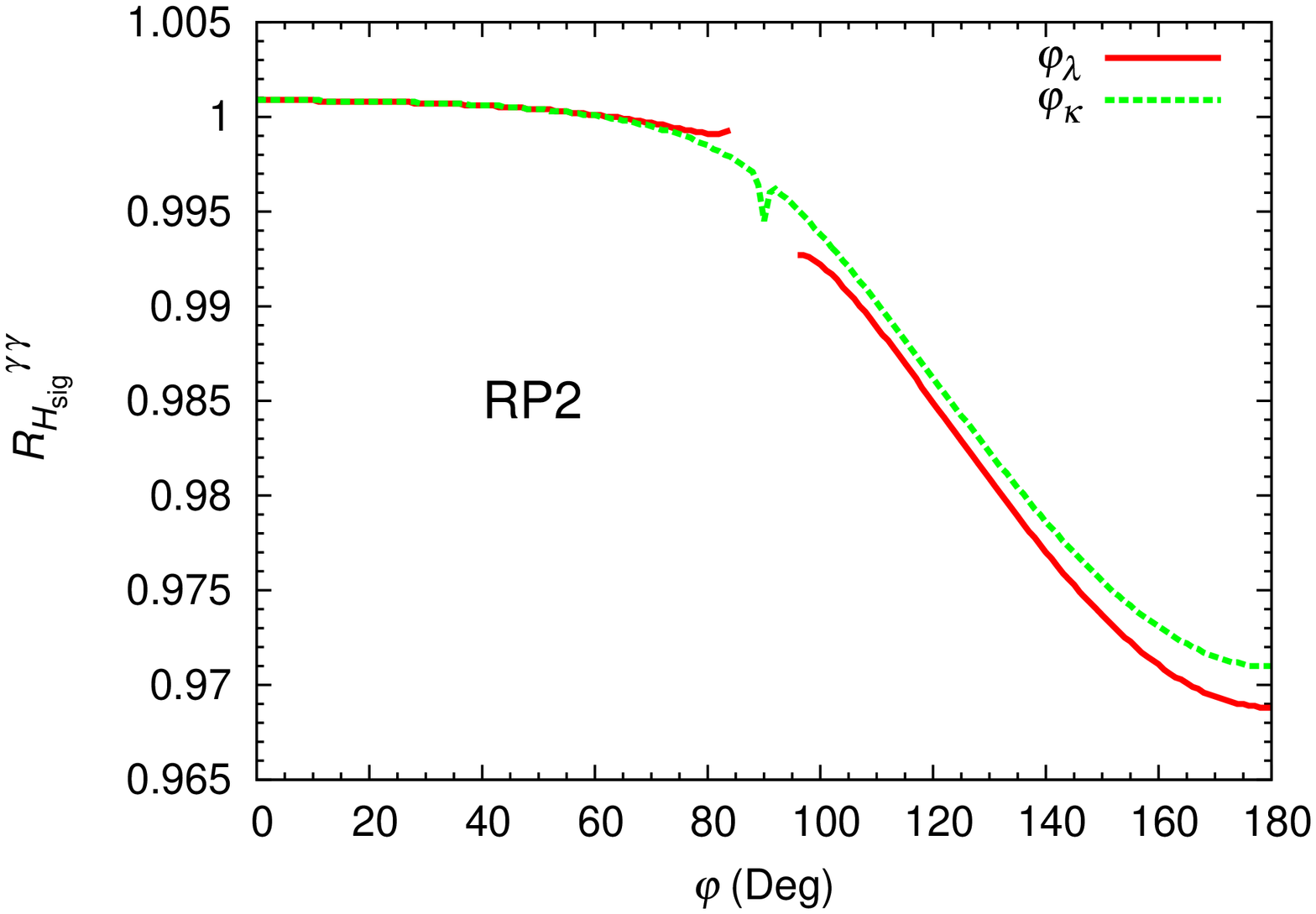}
}
\subfloat[]{%
\includegraphics*[angle=-90, scale=0.4]{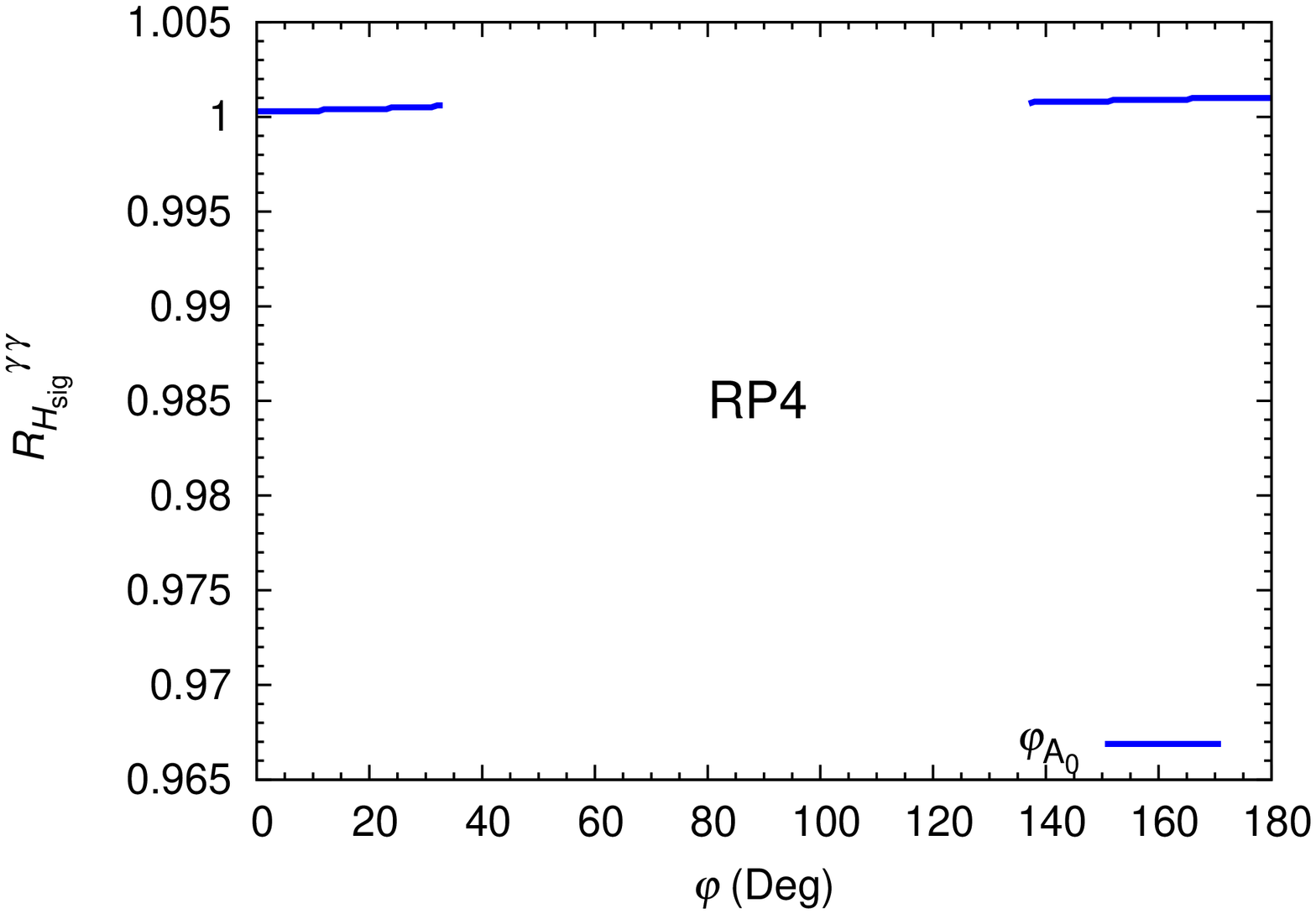}
}

\caption{Distributions of good points, $m_{H_{\rm sig}}$ and $R^{\gamma\gamma}_{H_{\rm sig}}$ for Scenario\,1, Case\,2.}
\label{fig:S1C2}
\end{figure} 

\noindent \underline{Case\,2:}  For large $\tan\beta$ in
Scenario\,1 the drop in the number of good points is slightly slower than in
Case\,1 with increasing \phlam/\phkap\ although it follows a similar
trend overall, as seen in
Fig.\,\ref{fig:S1C2}a. However, there is one notable distinction: the
number of surviving points in fact never falls to 0 for the entire
range of \phkap\ (although only about 100 points survive for
$\phkap>81^\circ$) while for a narrow range of \phlam\ not a single point
survives the imposed conditions. The reason for this will be explained below. A behavior
similar to the Case\,1 is also observed in Fig.\,\ref{fig:S1C2}b where
the number of good points falls to 0 for a relatively narrower range
of $\phtri$. The RPs for this Case are as follows. 
\begin{table}[h]
\begin{center}
\begin{tabular}{c|c|c|c|c}
 Point & $\lambda$ & $\kappa$ & $A_\lambda$\,(GeV) & $A_\kappa$\,(GeV) \\
\hline
RP1 & 0.086 & 0.12 & 1500 & -4000 \\
\hline
RP2 & 0.086 & 0.12 & 1500 & -3000\\
\hline
RP3 & 0.01 & 0.3 & 3000 & -4000\\
\hline
RP4 & 0.095 & 0.12 & 1500 & -1000\\
\end{tabular}
\end{center}
\end{table}

The sudden drop in the number of good points is, once again, largely driven by 
\mhsig\ which falls gradually with increasing
\phlam\ and \phkap, as seen in Fig.\,\ref{fig:S1C2}c
for RP1. However, this particular RP is one of the few points for
which the entire range of \phkap\ is allowed since $m_{H_1}$ never hits the
imposed lower limit. Notice the small break in the line
corresponding to \phlam\ despite the
lower limit on $m_{H_1}$ being satisfied, which results from the
falling of the mass
of the lightest chargino, $\chi_1^\pm$, below the LEP limit
($m_{\chi_1^\pm}>94$\,GeV \cite{Beringer:1900zz}) for
this range of \phlam. This in turn causes the
number of surviving points to fall to 0 for some intermediate values
of \phlam, as seen above. Note that this does not happen
for \phkap\ as this phase does not enter the chargino mass matrix directly, contrary to \phlam\ (see Appendix A). 
Fig.\,\ref{fig:S1C2}d for RP3 shows that $m_{H_1}$ in this Case can reach
comparatively higher values than in Case\,1, due to larger
$\tan\beta$, but its variation with increasing \phtri\ follows the same trend.

\rhsiggg\ for RP2, shown in Fig.\,\ref{fig:S1C2}e, again follows the
same trend with increasing \phlam/\phkap\ as in Case\,1, except
that the line corresponding to \phlam\ has a relatively small
break (which by contrast is due to the violation of the LEP constraint
on $m_{\chi_1^\pm}$ here, as noted above). On the other hand, the line corresponding to \phkap, while being
continuous, has a small kink around $\phkap=90^\circ$, where the mass
of the lightest singlet-like neutralino, $\chi_1^0$, becomes
small enough to kinematically allow decay of $H_1$ into its pair. This
causes a small drop in the BRs of $H_1$ into all SM particles. Note that, besides \phlam, \phkap\ also enters
the neutralino mass matrix directly, as opposed to the chargino mass
matrix. Fig.\,\ref{fig:S1C2}f for RP4 of this Case shows a slightly larger (although still
negligible) enhancement in \rhsiggg\ with increasing
\phtri\ as compared to that in the Case\,1. \\
 
We should mention here that $H_2$ and $H_3$ are always very heavy,
$\sim$1\,TeV and $\sim$2\,TeV, respectively, for both the cases
of this Scenario. Moreover, \rhsigzz\ follows the same trend with the
variation in CPV phases as \rhsiggg\ and is in fact always almost
equal to it. Some important numbers corresponding to these cases and to the four RPs in each of them are provided in Table\,\ref{tab:S1}. 

\subsection{Scenario\,2:}

\begin{figure}[p]
\centering
\subfloat[]{%
\includegraphics*[angle=-90, scale=0.4]{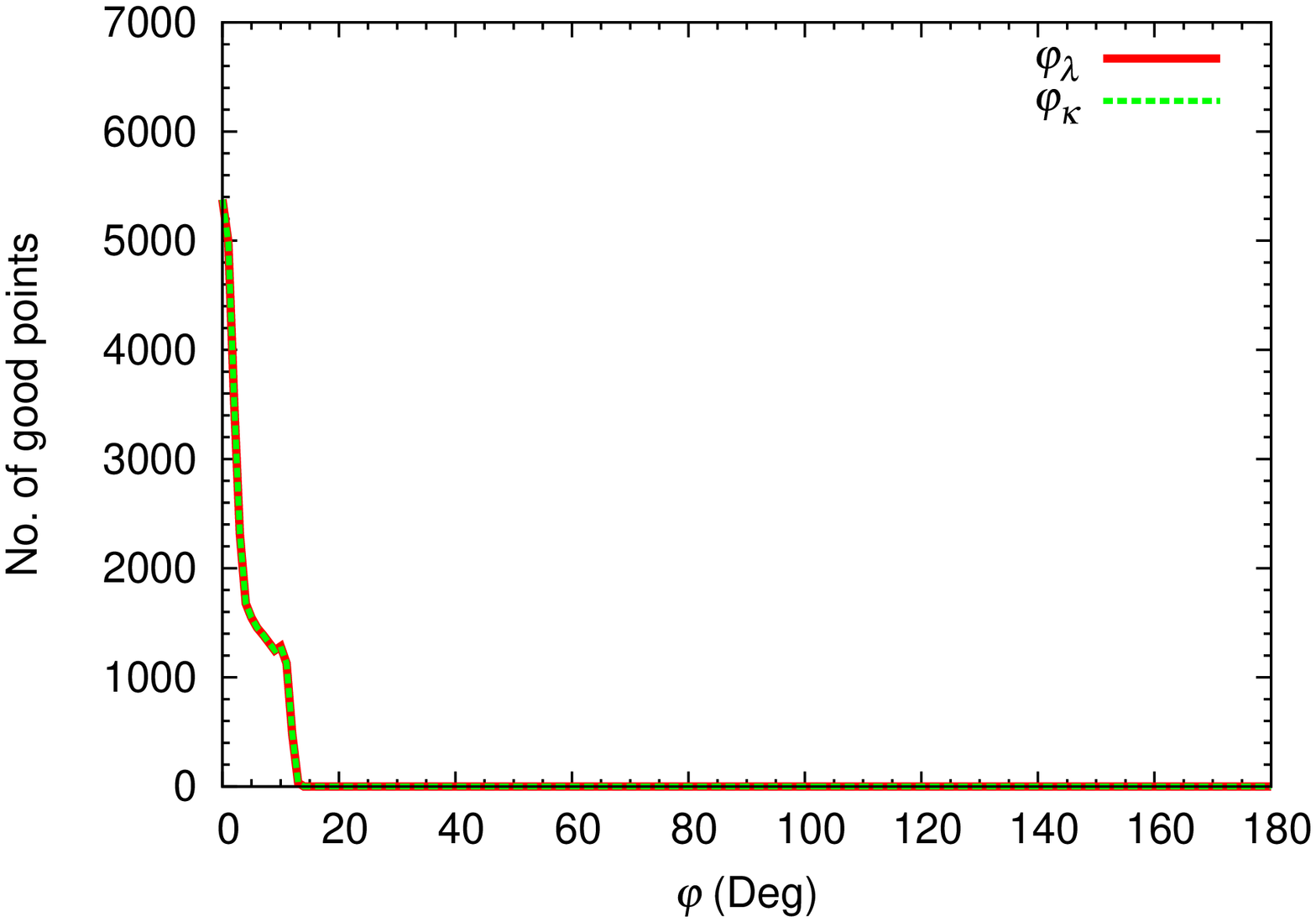}
}
\subfloat[]{%
\includegraphics*[angle=-90,scale=0.4]{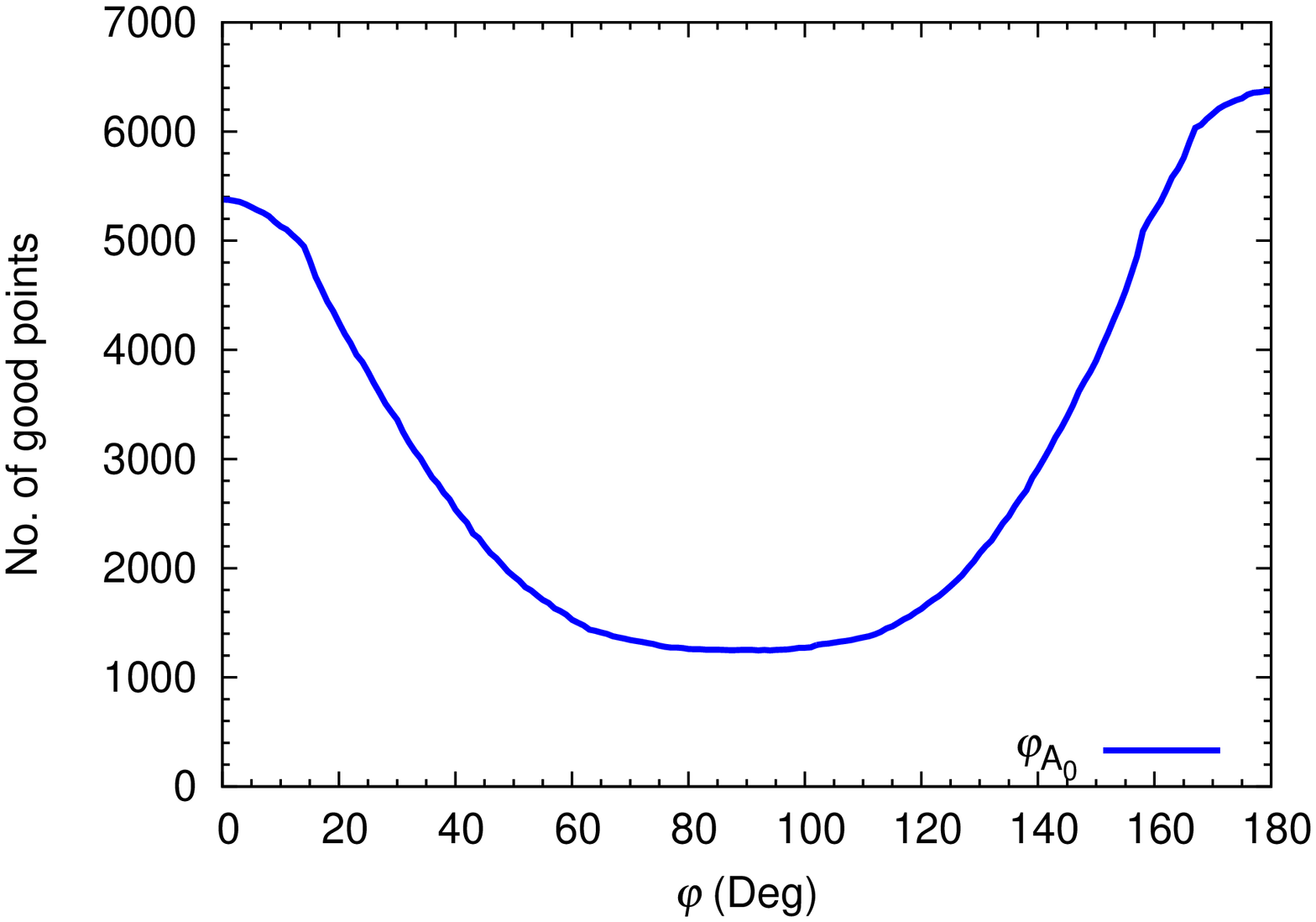}
}

\centering
\subfloat[]{%
\includegraphics*[angle=-90, scale=0.4]{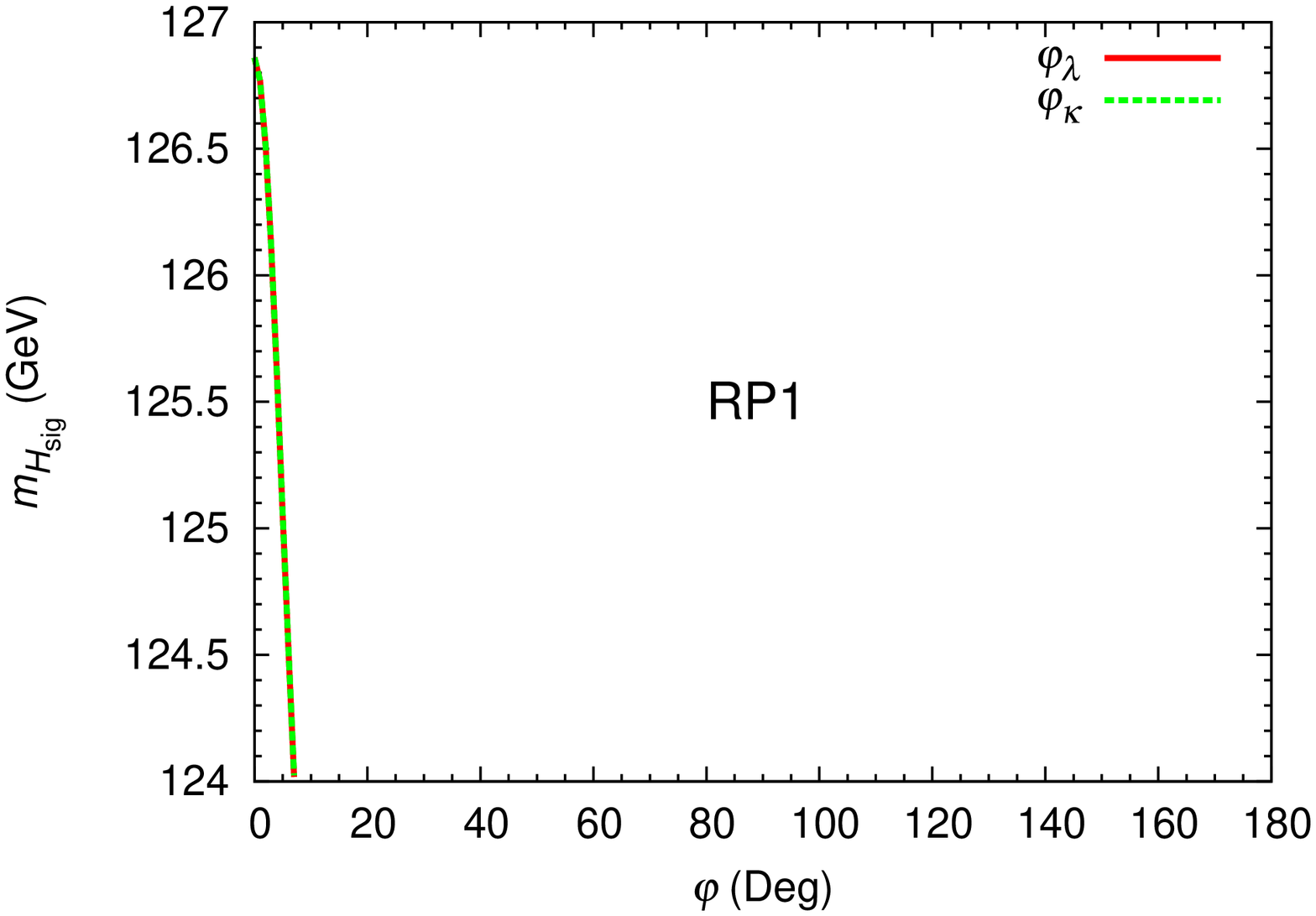}
}
\subfloat[]{%
\includegraphics*[angle=-90, scale=0.4]{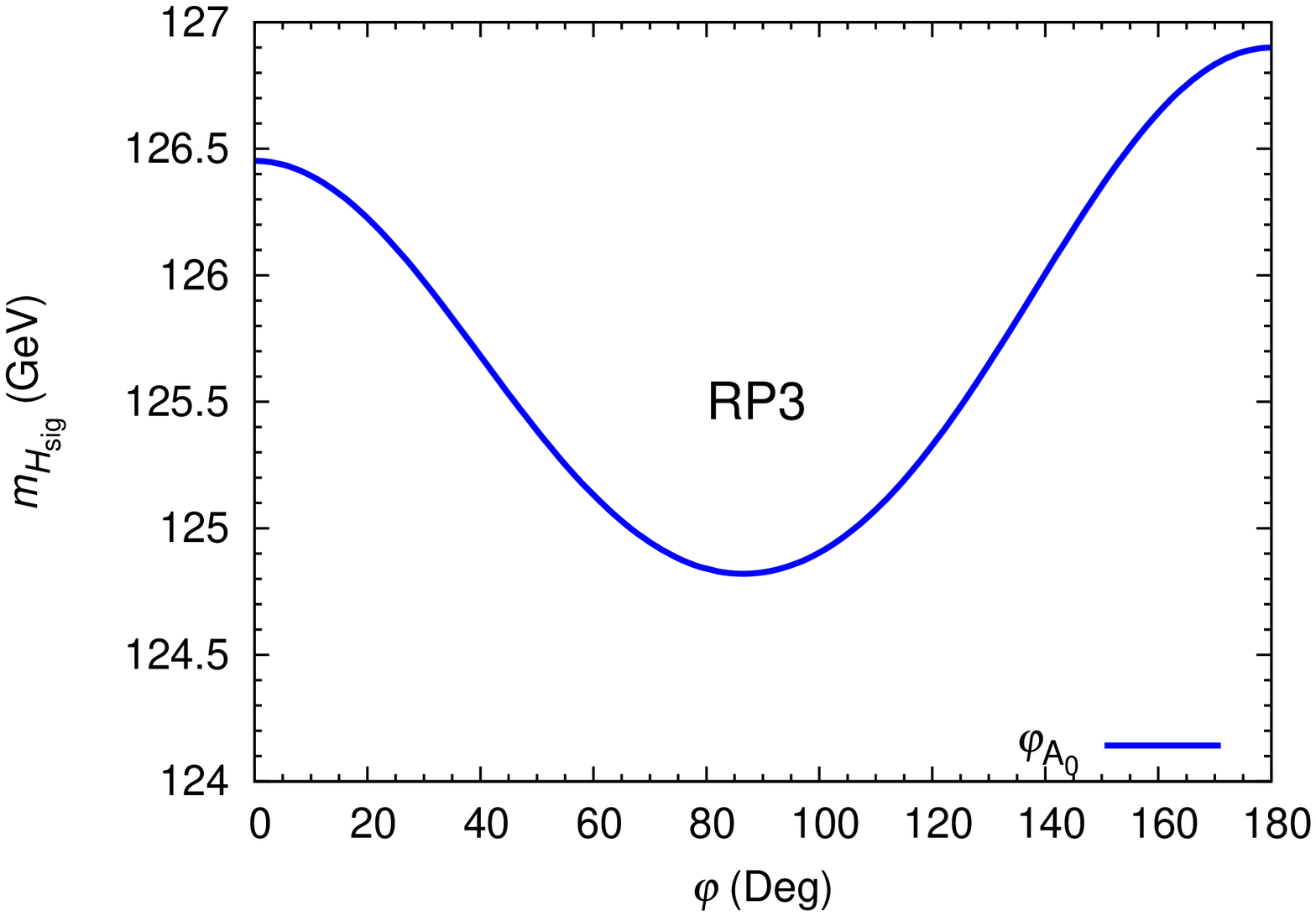}
}

\centering
\subfloat[]{%
\includegraphics*[angle=-90, scale=0.4]{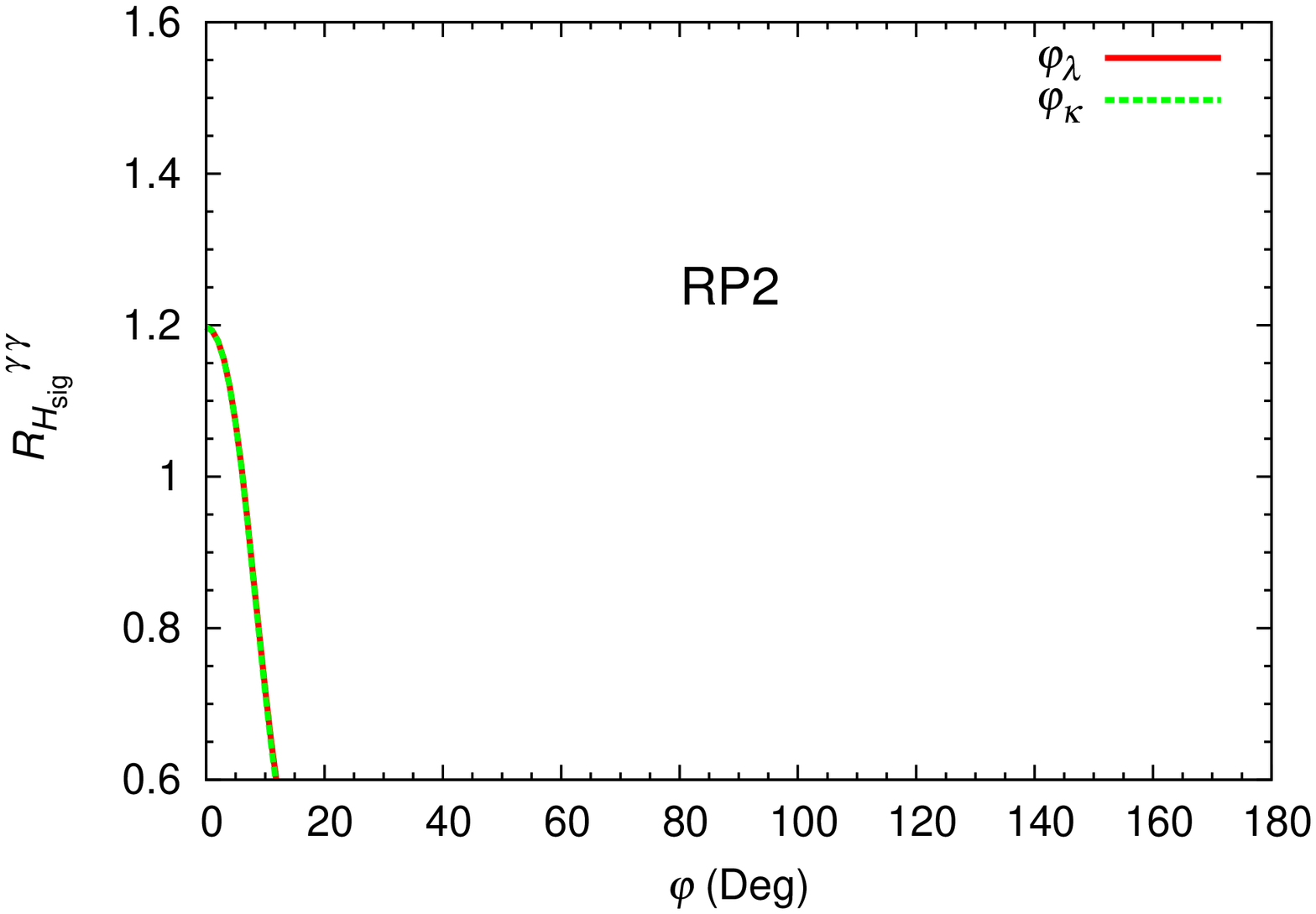}
}
\subfloat[]{%
\includegraphics*[angle=-90, scale=0.4]{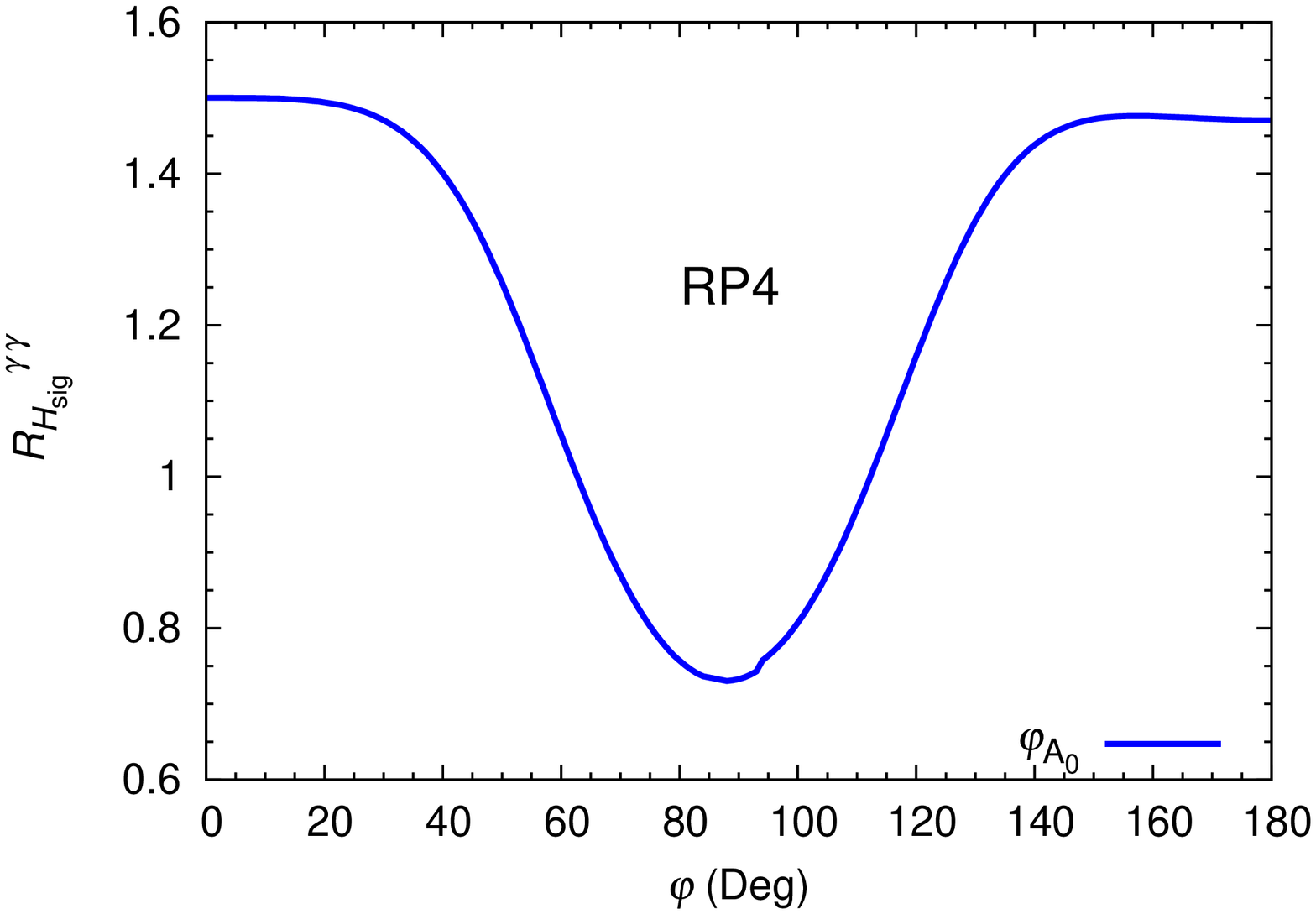}
}

\caption{Distributions of good points, $m_{H_{\rm sig}}$ and $R^{\gamma\gamma}_{H_{\rm sig}}$ for Scenario\,2, Case\,1.}
\label{fig:S2C1}
\end{figure}

\noindent \underline{Case\,1:} In Fig.\,\ref{fig:S2C1}a we show the
number of good points against the phases \phlam\ and \phkap\ for this Case. We see that the number of
surviving points for $\phlam/\phkap=0^\circ$ is much smaller compared to that in the two cases of
Scenario\,1. Moreover, there is comparatively an even steeper drop 
in the number of surviving points with increasing \phlam\ and
\phkap. In fact, the number of surviving points
falls to 0 for $\phlam/\phkap$ as low as $14^\circ$, owing again to
the sensitivity of $m_{H_2}$ to these phases, and it stays 0 for their larger
values. The number of surviving points, on the other hand,
never falls below $\sim$1250 for the entire range of \phtri, as seen
in Fig.\,\ref{fig:S2C1}b. In fact, the number of good points falls
slowly with increasing \phtri, becomes almost constant with the latter
between $\sim70^\circ$ and $\sim110^\circ$, and then starts rising
again so that for $\phtri=180^\circ$ it is even larger than in the CPC case.
The particulars of the RPs for this Case are given below. 
\begin{table}[h]
\begin{center}
\begin{tabular}{c|c|c|c|c}
 Point & $\lambda$ & $\kappa$ & $A_\lambda$\,(GeV) & $A_\kappa$\,(GeV) \\
\hline
RP1 & 0.558 & 0.36 & 187 & -228\\
\hline
RP2 & 0.511 & 0.34 & 173 & -222 \\
\hline
RP3 & 0.516 & 0.33 & 187 & -200 \\
\hline
RP4 & 0.553 & 0.36 & 173 & -217 \\
\end{tabular}
\end{center}
\end{table}

\begin{table}[t]
\begin{center}
\begin{tabular}{|l|c|c|}
\hline
Scenario & 1,\,Case\,1 & 1,\,Case\,2  \\
\hline
\hline
Points scanned for each $\phi_\kappa$ or $\phi_\lambda$ or
$\phi_{A_0}$ &\multicolumn{2}{|c|}{40000} \\
\hline
Points surviving for $\phi_\kappa =\phi_\lambda = \phi_{A_0}  = 0$ &
39961 & 39954   \\
\hline 
\hline
Min. points surviving, with $\phi_\kappa$ & 0,\,78-144 & 100,\,82-180 \\
Max. points surviving, with $\phi_\kappa$ & 39961,\,0-2
& 39954,\,0-2\\
Min. points surviving, with $\phi_\lambda$ & 0,\,76-155& 0,\,85-95\\
Max. points surviving, with $\phi_\lambda$ & 39961,\,0-2 &
39954,\,0-2 \\
\hline

$m_{h_{\rm sig}}$ for RP1 with $\phi_\kappa =\phi_\lambda = 0$ & 124.7990 & 125.1936\\
\rhsiggg\ for RP2 with $\phi_\kappa =\phi_\lambda = 0$ & 1.0003 & 1.0010 \\
\hline

Min. $m_{h_{\rm sig}}$ obtained for RP1, with $\phi_\kappa$ &
124.0159,\,160 & 124.2068,\,180\\
Min. \rhsiggg\ obtained for RP2, with $\phi_\kappa$ &
0.9801,\,162 & 0.9710,\,180 \\
\hline

Min. $m_{h_{\rm sig}}$ obtained for RP1, with $\phi_\lambda$ &
124.0058,\,39  & 124.1254,\,180 \\
Min. \rhsiggg\ obtained for RP2, with $\phi_\lambda$ &
0.9792,\,180 & 0.9688,\,180 \\
\hline 
\hline

Min. points surviving, with $\phi_{A_0}$ & 0,\,27-142 & 0,\,38-133 \\
Max. points surviving, with $\phi_{A_0}$ & 39987,\,171-180 &
39973,\,171-180 \\
\hline
$m_{h_{\rm sig}}$ for RP3 with $\phi_{A_0}$ = 0 & 124.9199 & 125.9711 \\
\rhsiggg\ for RP4 with $\phi_{A_0} = 0$ & 1.0014 & 1.0003 \\
\hline
 Min. $m_{h_{\rm sig}}$ obtained for RP3, with $\phi_{A_0}$ &
 124.0502,\,26 & 124.0737,\,134\\
 Max. $m_{h_{\rm sig}}$ obtained for RP3, with $\phi_{A_0}$ &
 125,5270,\, 180 & 126.5914,\,180\\
 Max. \rhsiggg\ obtained for RP4, with $\phi_{A_0}$ &
1.0017,\,180 & 1.0010,\,180 \\
\hline 

\end{tabular}
\caption{Scan results for Scenario\,1, Cases 1 and 2. All angles are in degrees.}
\label{tab:S1}
\end{center}
\end{table}


\mhsig\ in this Case can easily reach the defined upper limit of
127\,GeV when CP is conserved but falls very abruptly with increasing \phlam/\phkap, as shown
for RP1 in Fig.\,\ref{fig:S2C1}c.
The reason is that this Case corresponds to larger
values of $\lambda$ and $\kappa$ compared to any other Case discussed
here and, consequently, the
dependence on their phases is more pronounced. In
Fig.\,\ref{fig:S2C1}d we show the dependence of $m_{H_2}$ on
\phtri\ for RP3. While \mhsig\
here stays above the imposed lower limit for all
values of \phtri, in contrast with what was observed for Scenario\,1, its overall behavior is quite
similar. \mhsig\ again falls continuously with
increasing \phtri\ until some intermediate value of the latter
and then starts rising, reaching a value for $\phtri=180^\circ$
that is larger than the value for the CPC
case. Note that in this Case also it is possible to find
points with \mhsig\ falling as sharply and reaching the imposed lower
limit with not too large \phtri, as
was seen in Scenario\,1. However, our selected RP3 demonstrates well the possibility
of the entire range of \phtri\ being allowed, which is precluded in Scenario\,1.  

Fig.\,\ref{fig:S2C1}e for RP2 shows that while \rhsiggg\ can be much
higher than the SM expectation for the CPC case, the drop in it is
very steep with increasing \phlam/\phkap. This is because the total
width of $H_2$ falls sharply owing again to the fact that $\lambda$ and 
$\kappa$ have fairly large absolute values. 
This in fact results in a slow rise in BR$(H_2\rightarrow
\gamma\gamma)$ compared to the CPC case. But since the partial width
of $H_2$ into $gg$ falls comparatively faster, it causes an overall drop in \rhsiggg.
Fig.\,\ref{fig:S2C1}f for RP4 shows an initially slow but eventually
sharp drop in \rhsiggg\ with increasing
\phtri\ until the latter reaches its intermediate values when \rhsiggg\
starts rising again. The main reason for the initial slow
drop with increasing \phtri\ is that while $\Gamma(H_2\rightarrow
gg)$ always keeps dropping slowly BR$(H_2\rightarrow \gamma\gamma)$ initially
stays almost constant but later starts falling also. Evidently, this
behavior is reversed after $\phtri=90^\circ$, when \rhsiggg\ starts rising
again. Finally, \rhsigzz\ in this Case is
always considerably lower than \rhsiggg\ (e.g., it is $\sim$0.7 for the
CPC case of RP2) but shows a similar behavior with
varying CPV phases. \\

\begin{figure}[p]
\centering
\subfloat[]{%
\includegraphics*[angle=-90, scale=0.4]{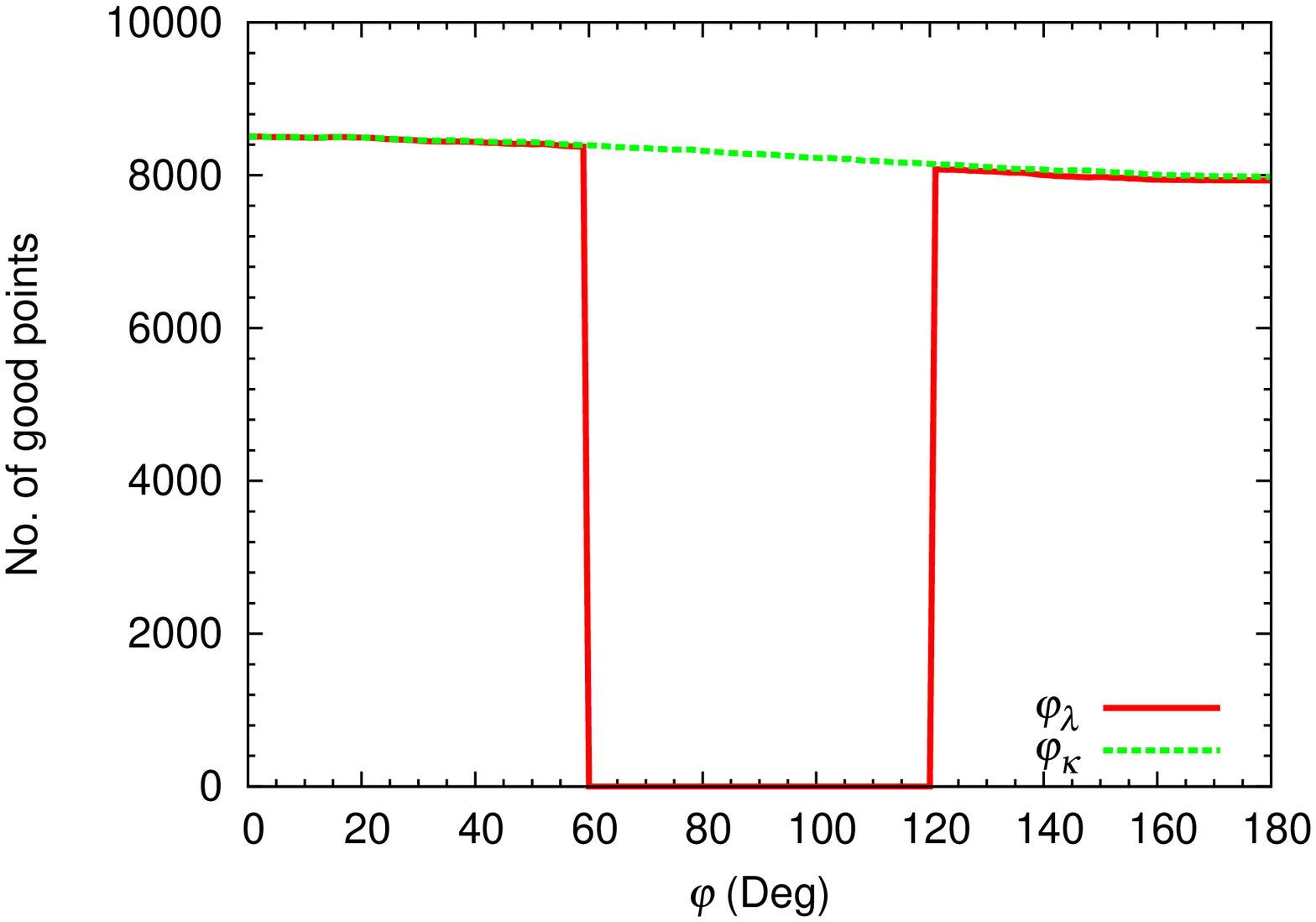}
}
\subfloat[]{%
\includegraphics*[angle=-90,scale=0.4]{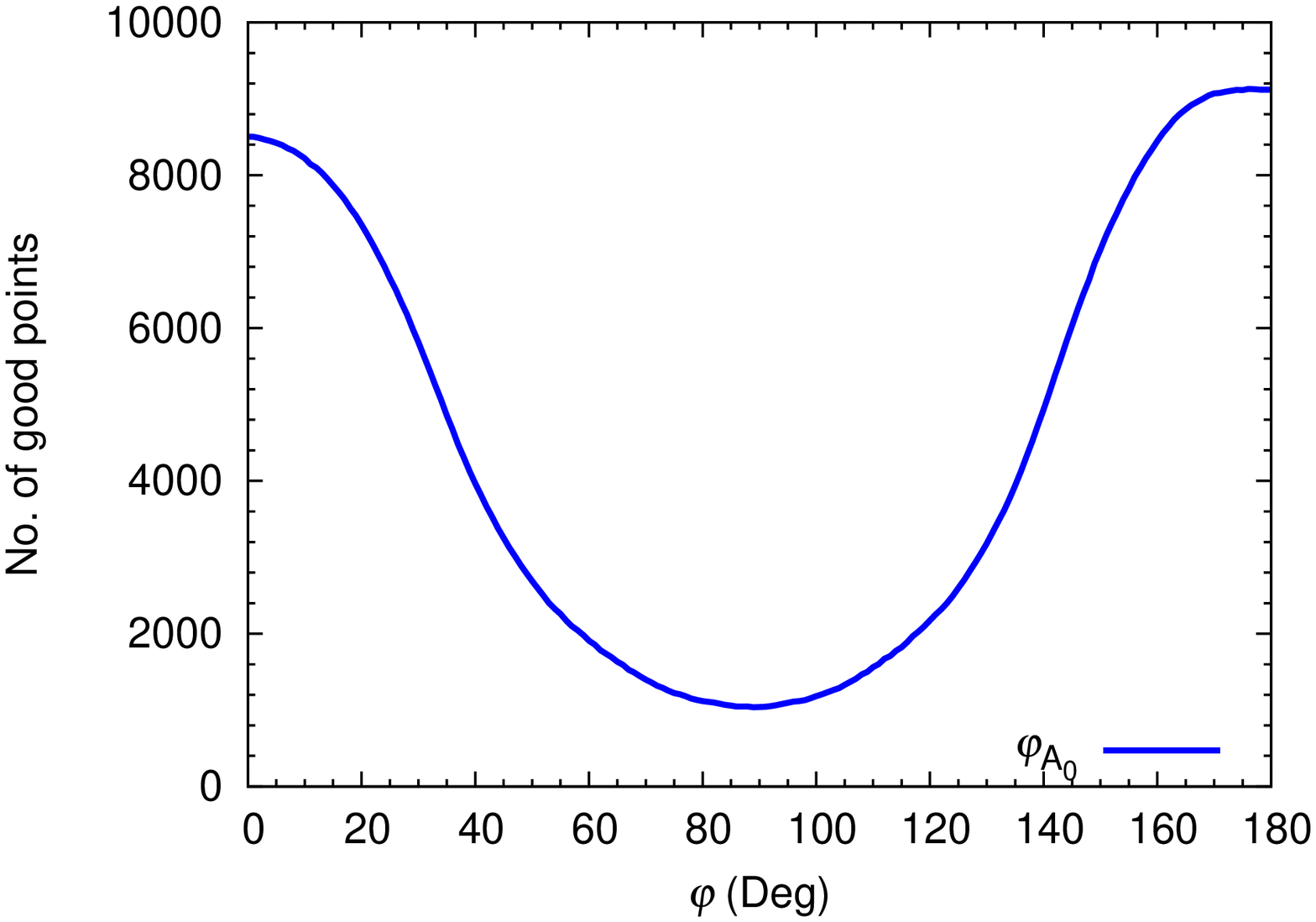}
}

\centering
\subfloat[]{%
\includegraphics*[angle=-90, scale=0.4]{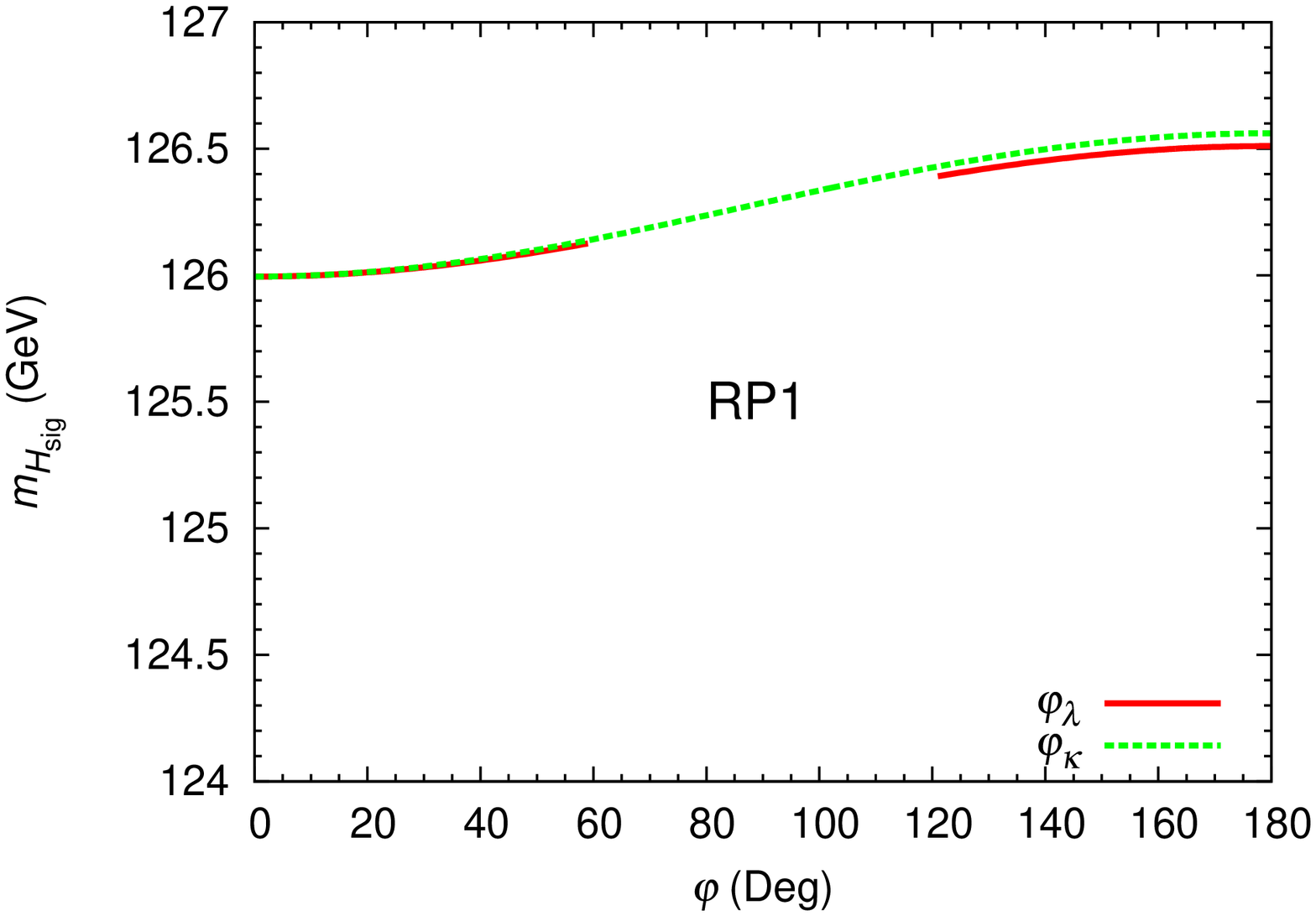}
}
\subfloat[]{%
\includegraphics*[angle=-90, scale=0.4]{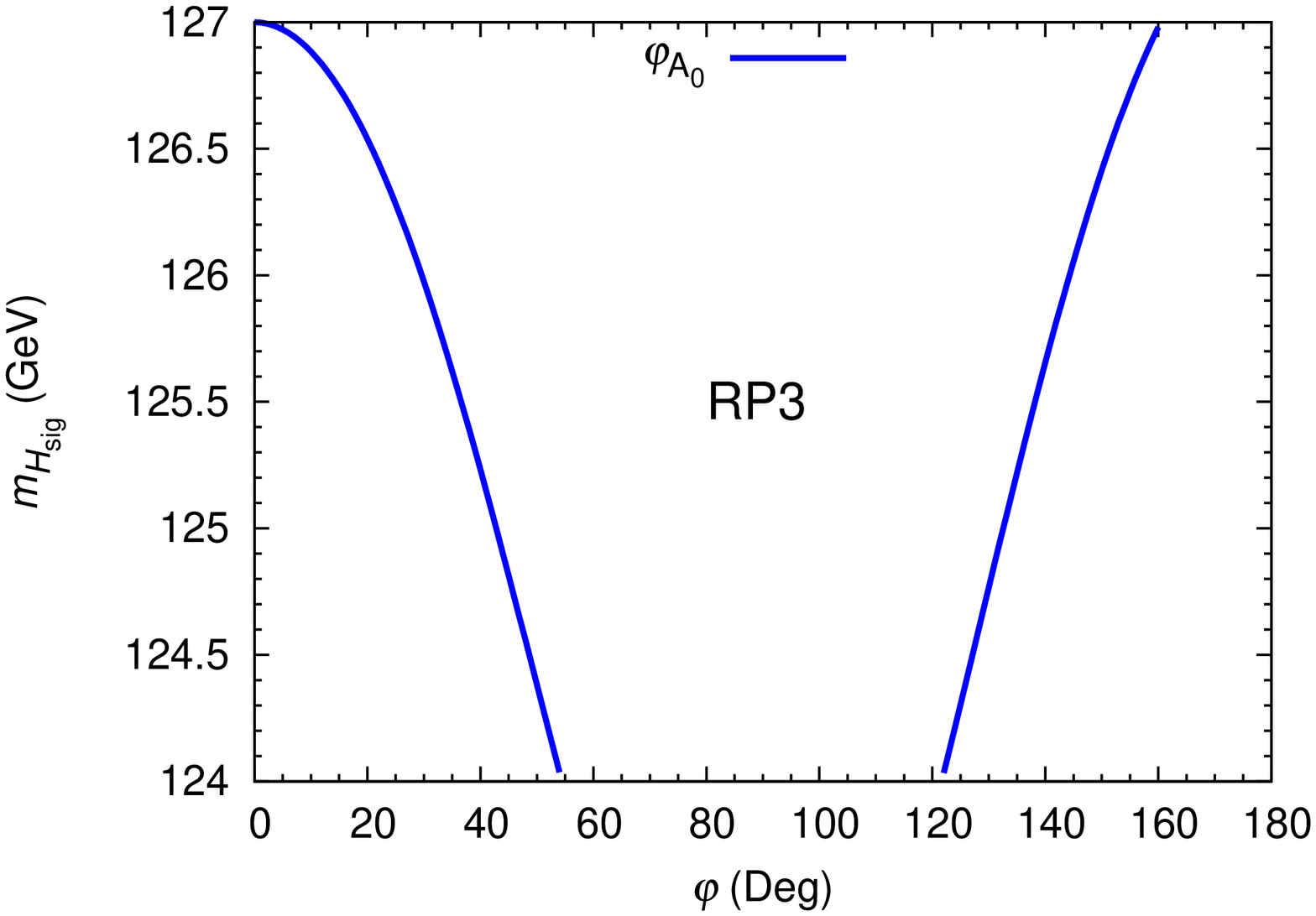}
}

\centering
\subfloat[]{%
\includegraphics*[angle=-90, scale=0.4]{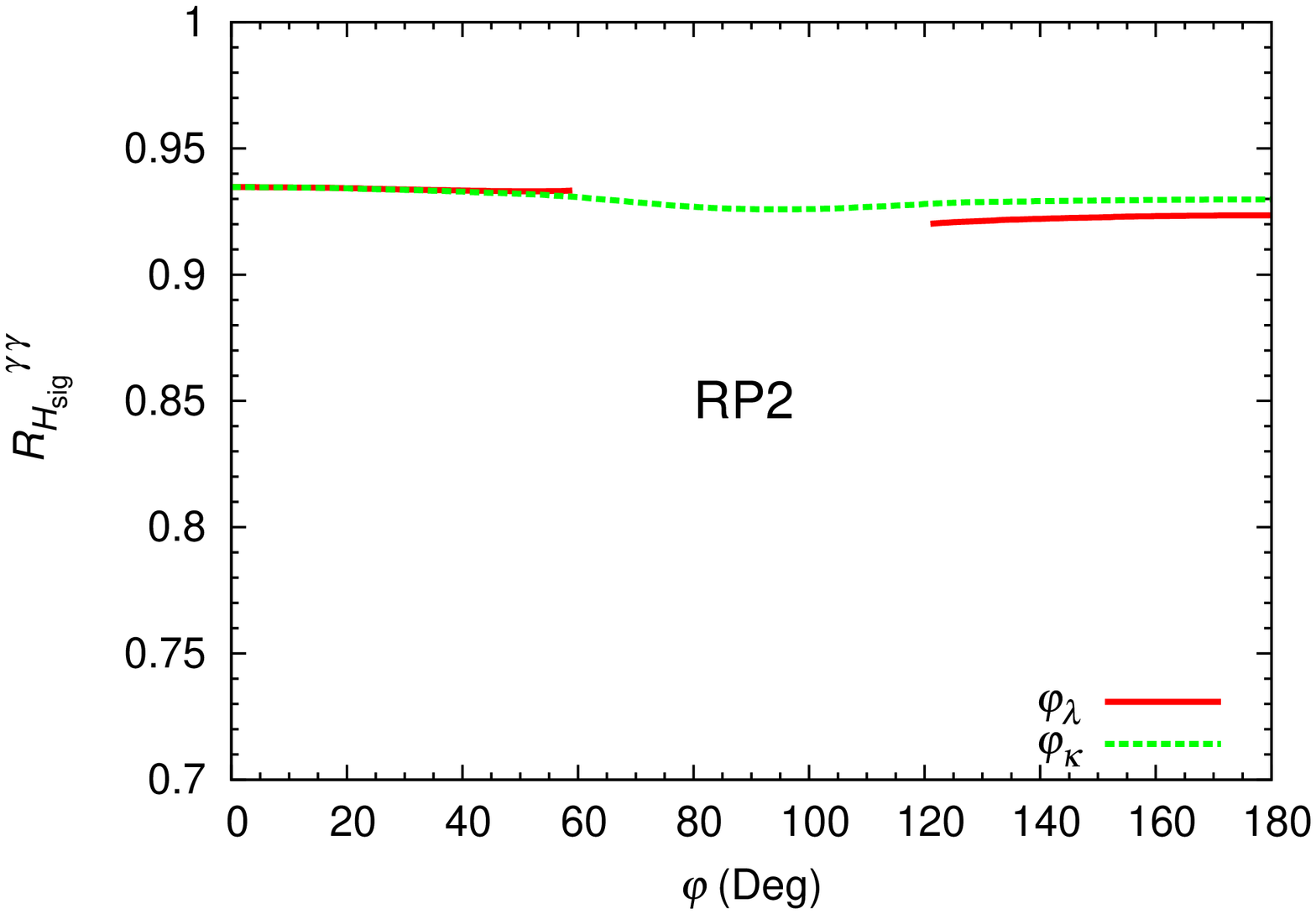}
}
\subfloat[]{%
\includegraphics*[angle=-90, scale=0.4]{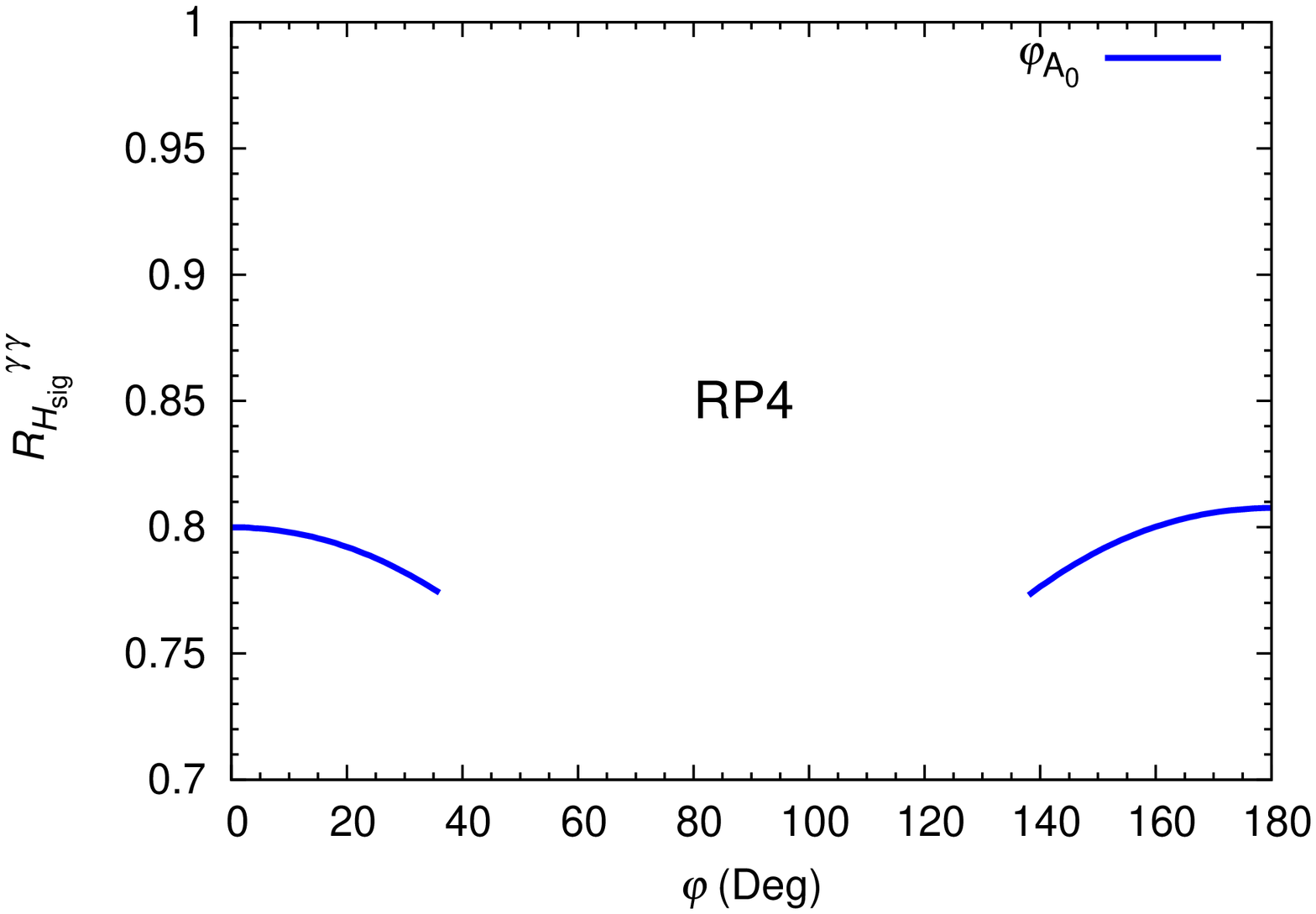}
}
\caption{Distributions of good points, $m_{H_{\rm sig}}$ and $R^{\gamma\gamma}_{H_{\rm sig}}$ for Scenario\,2, Case\,2.}
\label{fig:S2C2}
\end{figure}

\noindent \underline{Case\,2:} The composition of $H_2$ in this Case
is closer to that of $H_1$ in Scenario\,1 than to $H_2$ in Case\,1 of this
Scenario above (due to smaller $\lambda$ and consequently smaller
singlet component) state. However, Fig.\,\ref{fig:S2C2}a shows a somewhat different behavior 
 from Scenario\,1, as the number of good points, though comparatively
 much smaller for the CPC case, falls very
 slightly over the entire range of \phkap. For \phlam\ between
 $60^\circ$ and $120^\circ$, the number of points is reduced to
 0 due again to the violation of the LEP limit on
 $m_{\chi_1^\pm}$. The behavior of the number of good points with
 increasing \phtri, on the other hand, is very similar to the Case\,1 of
 this Scenario, as seen in Fig.\,\ref{fig:S2C2}b since for a large number of points \mhsig\ stays within the
 imposed limits for the entire range of \phtri. Below we give the values
of other Higgs sector parameters for the four RPs of this Case.

\begin{table}[h]
\begin{center}
\begin{tabular}{c|c|c|c|c}
 Point & $\lambda$ & $\kappa$ & $A_\lambda$\,(GeV) & $A_\kappa$\,(GeV) \\
\hline
RP1 & 0.043 & 0.015 & 200 & -160 \\
\hline
RP2 & 0.047 & 0.011 & 600 & -140 \\
\hline
 RP3 & 0.044 & 0.017 & 289 & -180\\
\hline
RP4 & 0.036 & 0.014 & 422 & -173\\
\end{tabular}
\end{center}
\end{table}

Fig.\,\ref{fig:S2C2}c for RP1 shows that, contrary to all the cases
discussed so far, \mhsig\ rises,
albeit slowly, with increasing \phlam/\phkap. The reason is the 
singlet-doublet mixing which causes, in turn, the mass of $S_R$-like $H_1$ to
fall. This is in somewhat analogy with Scenario\,1 wherein also the mass of
$H_2$, which is $S_R$-like instead, rises while that of the
$H_{dR}$-like $H_1$ falls with increasing 
amount of CP-violation. The dependence of \mhsig\ on
\phtri\ for RP3, shown in Fig.\,\ref{fig:S2C2}c,
is still similar to what has been observed so far. Note again that while for this
particular RP \mhsig\ touches the allowed upper limit for the CPC
case and drops sharply to the lower limit, thus excluding a wide range
of \phtri, points similar to RP3 of the Case\,1 above (with the entire
range of \phtri\ allowed) are also available.   

In Fig.\,\ref{fig:S2C2}e we show the variation
in \rhsiggg\ with \phlam\ and \phkap\ for RP2 of this Case. 
When CP is conserved in this Case \rhsiggg\ is generally slightly
lower than 1 (SM expectation) but still lies well within the observed
range taking into account the experimental uncertainties, reported by
the CMS collaboration \cite{LHCresults2cms}. The behavior of \rhsiggg, 
which falls slowly with increasing \phlam/\phkap, is, however, remarkably similar to that
observed in Scenario\,1 (notice the relatively compressed scale
of the y-axis in the figure). With increasing but small (very large) \phtri\, on the other
hand, \rhsiggg\ falls (rises) a little faster than what has been noted
for earlier cases, as seen in Fig.\,\ref{fig:S2C2}f for RP4. This is due to the fact that,
conversely to the earlier cases, with increasing (but small) \phtri,
the dominant BR$(H_2\rightarrow b\bar{b})$ increases gradually while
BR$(H_2\rightarrow \gamma\gamma)$ itself falls very slowly, and vice
versa for very large \phtri. Once again, \rhsigzz\ is very
close to \rhsiggg\ for RP2 and RP4 in this Case also and follows a similar
trend in variation with an increase in any of the three CPV phases. \\

Some particular values corresponding to the benchmark points for the
two cases of this Scenario are given in Table\,\ref{tab:S2S3}. 

\subsection{Scenario\,3:}

\begin{figure}[p]
\centering
\subfloat[]{%
\includegraphics*[angle=-90, scale=0.4]{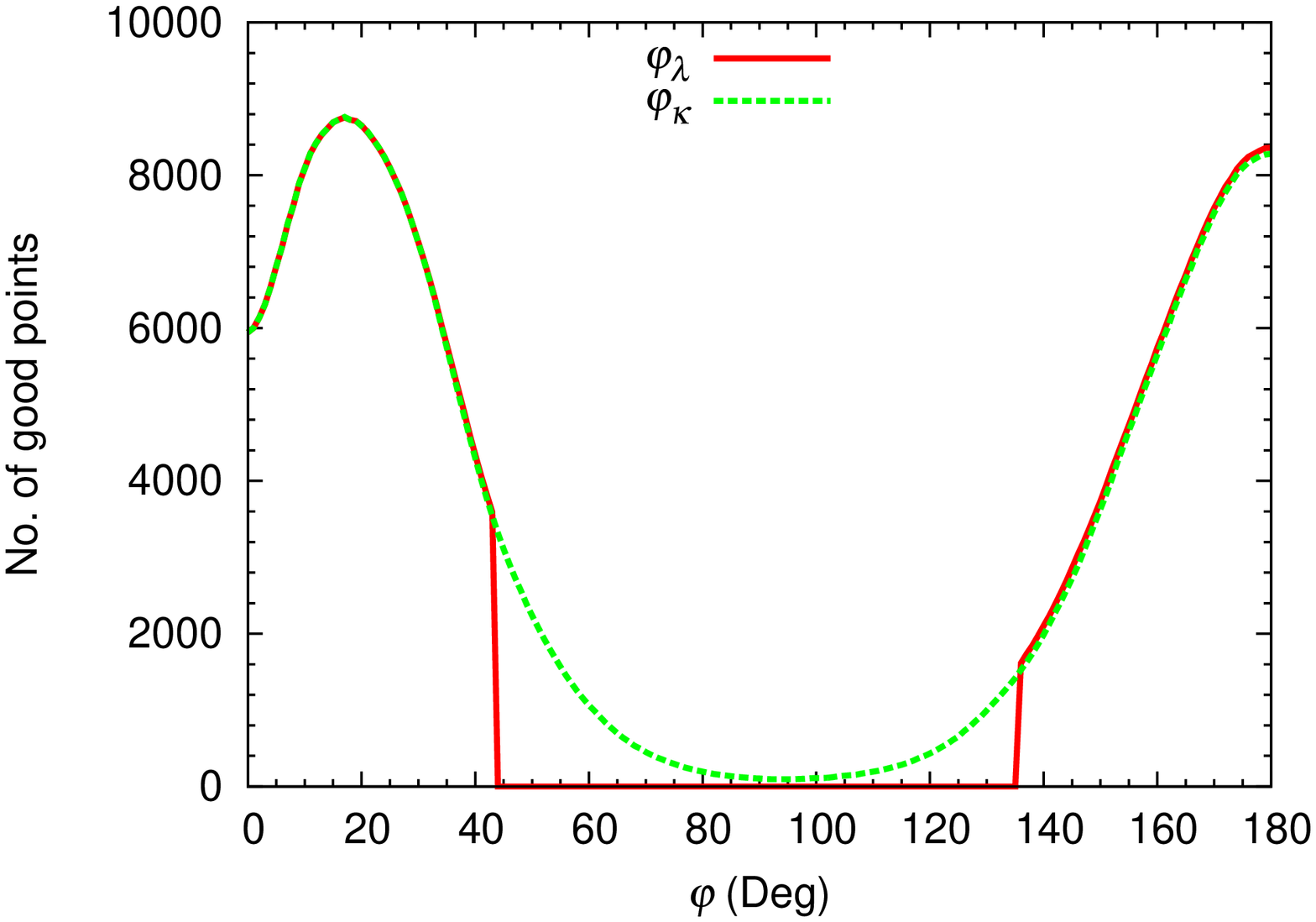}
}
\subfloat[]{%
\includegraphics*[angle=-90,scale=0.4]{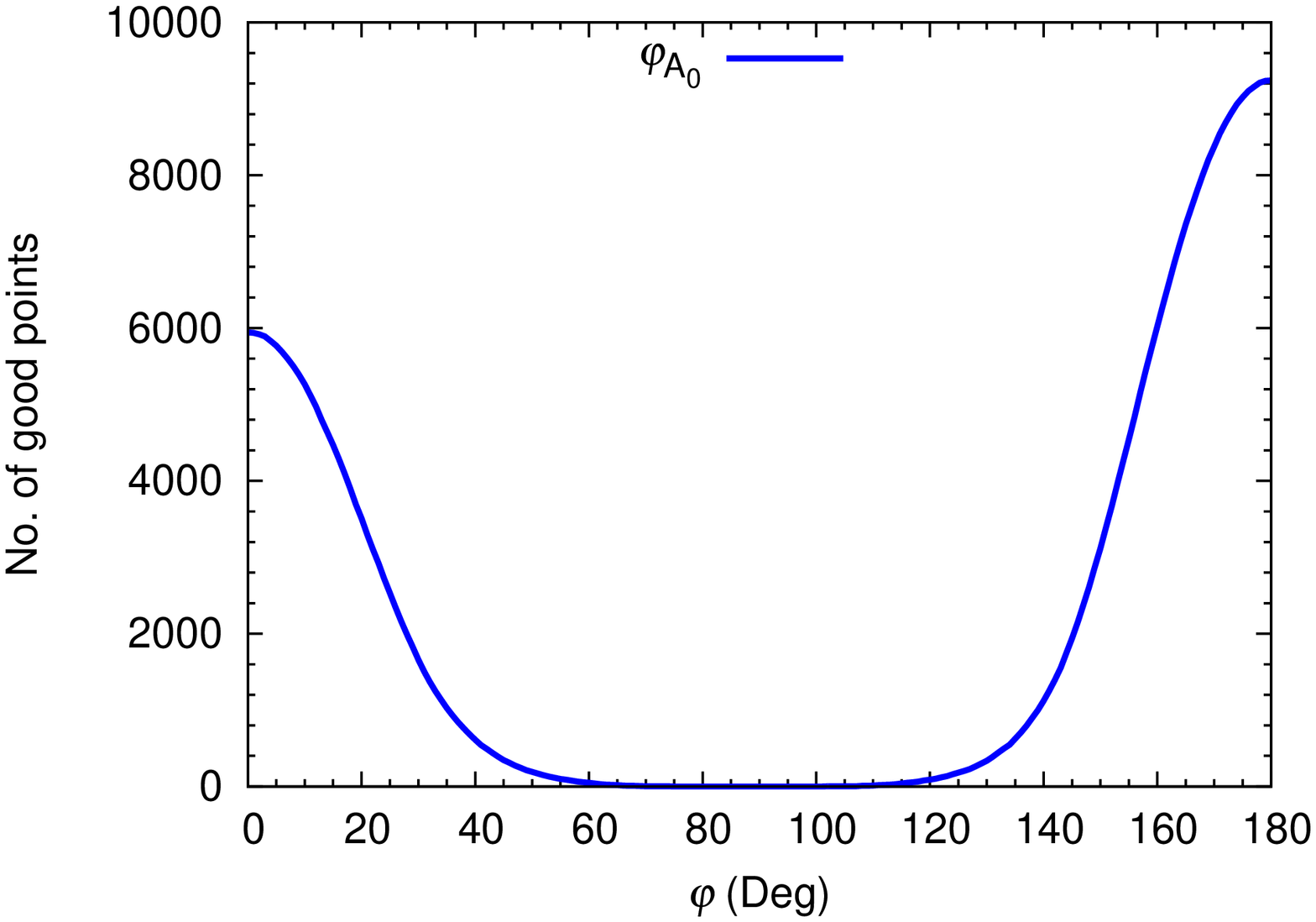}
}

\centering
\subfloat[]{%
\includegraphics*[angle=-90, scale=0.4]{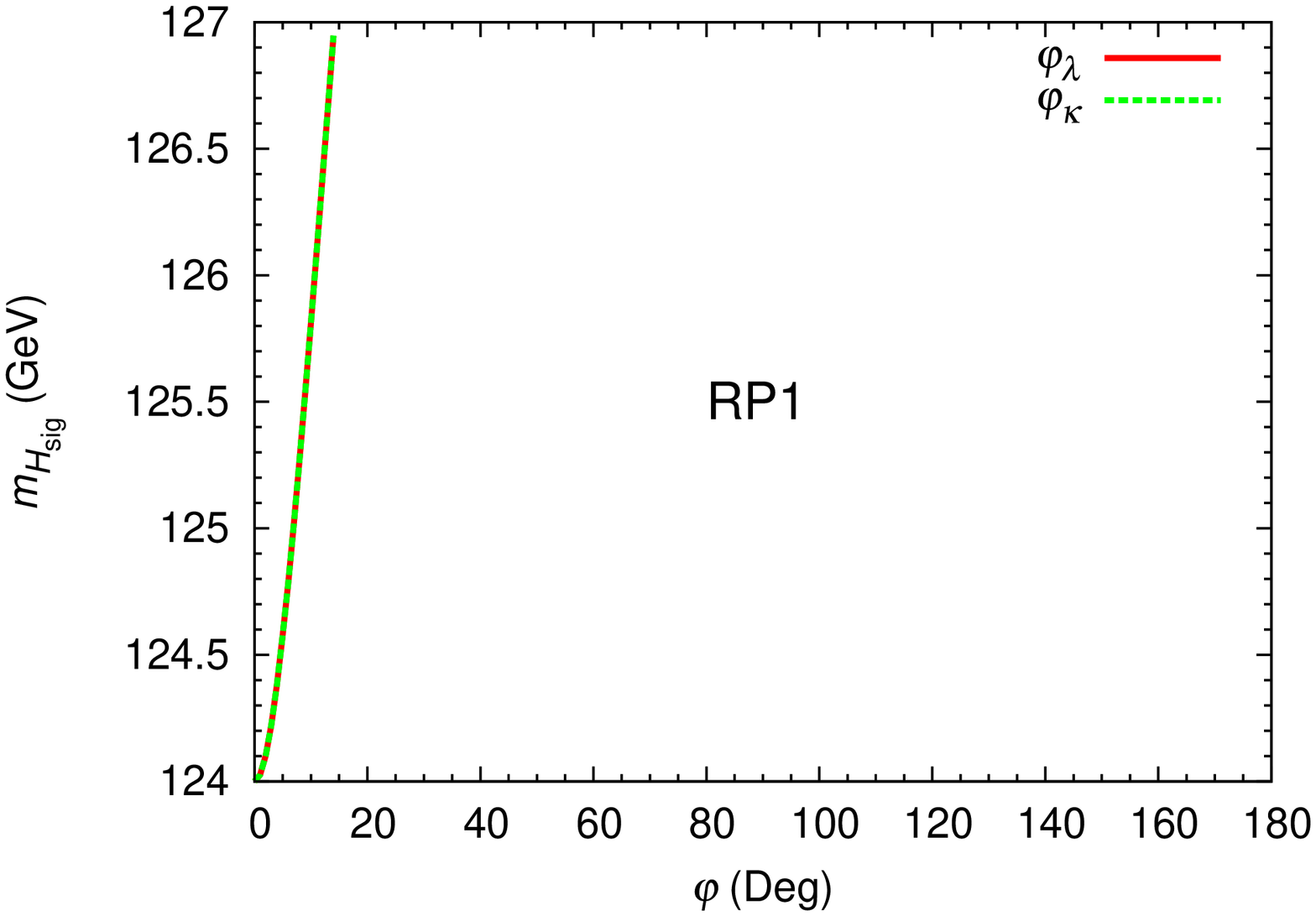}
}
\subfloat[]{%
\includegraphics*[angle=-90, scale=0.4]{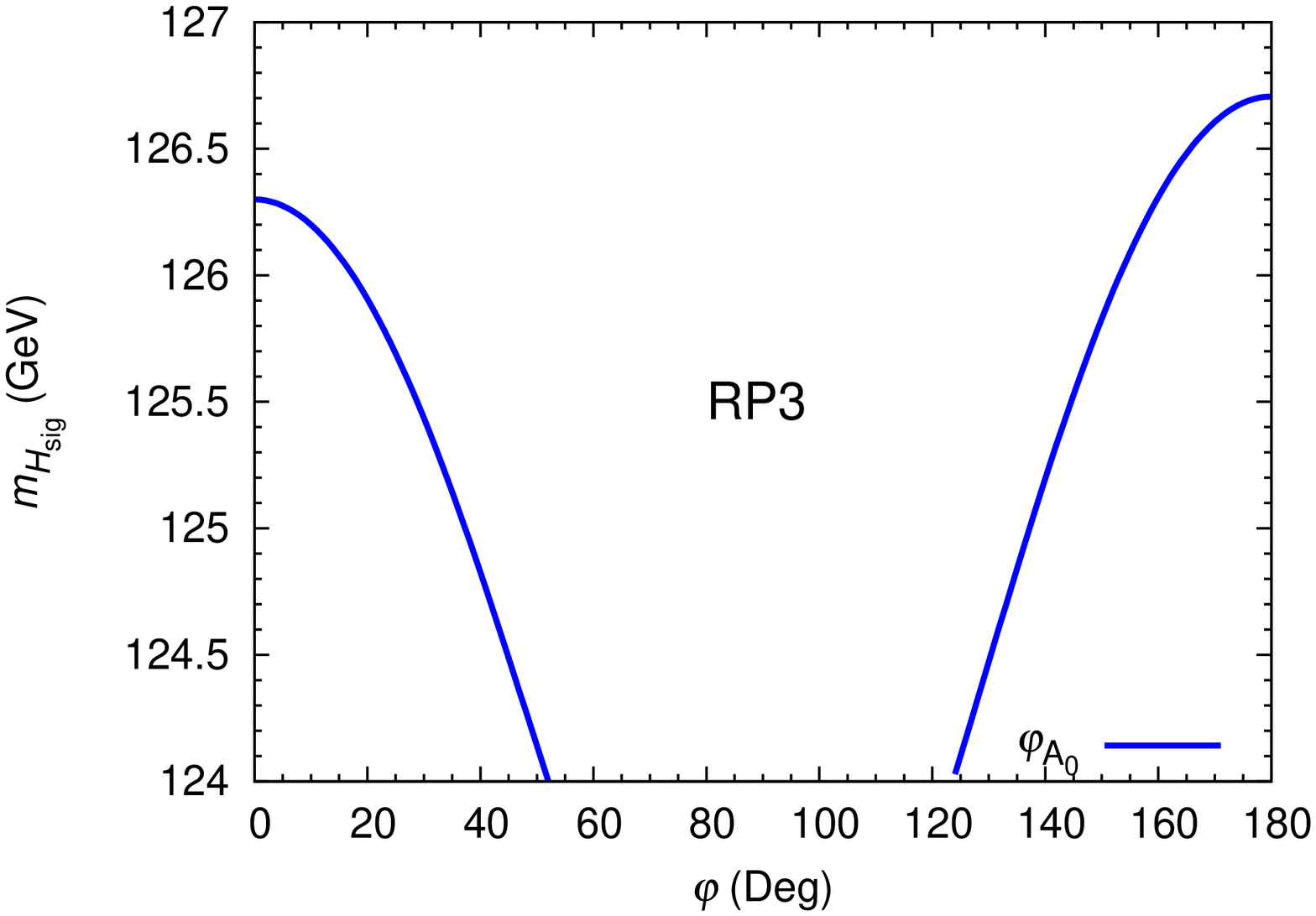}
}

\centering
\subfloat[]{%
\includegraphics*[angle=-90, scale=0.4]{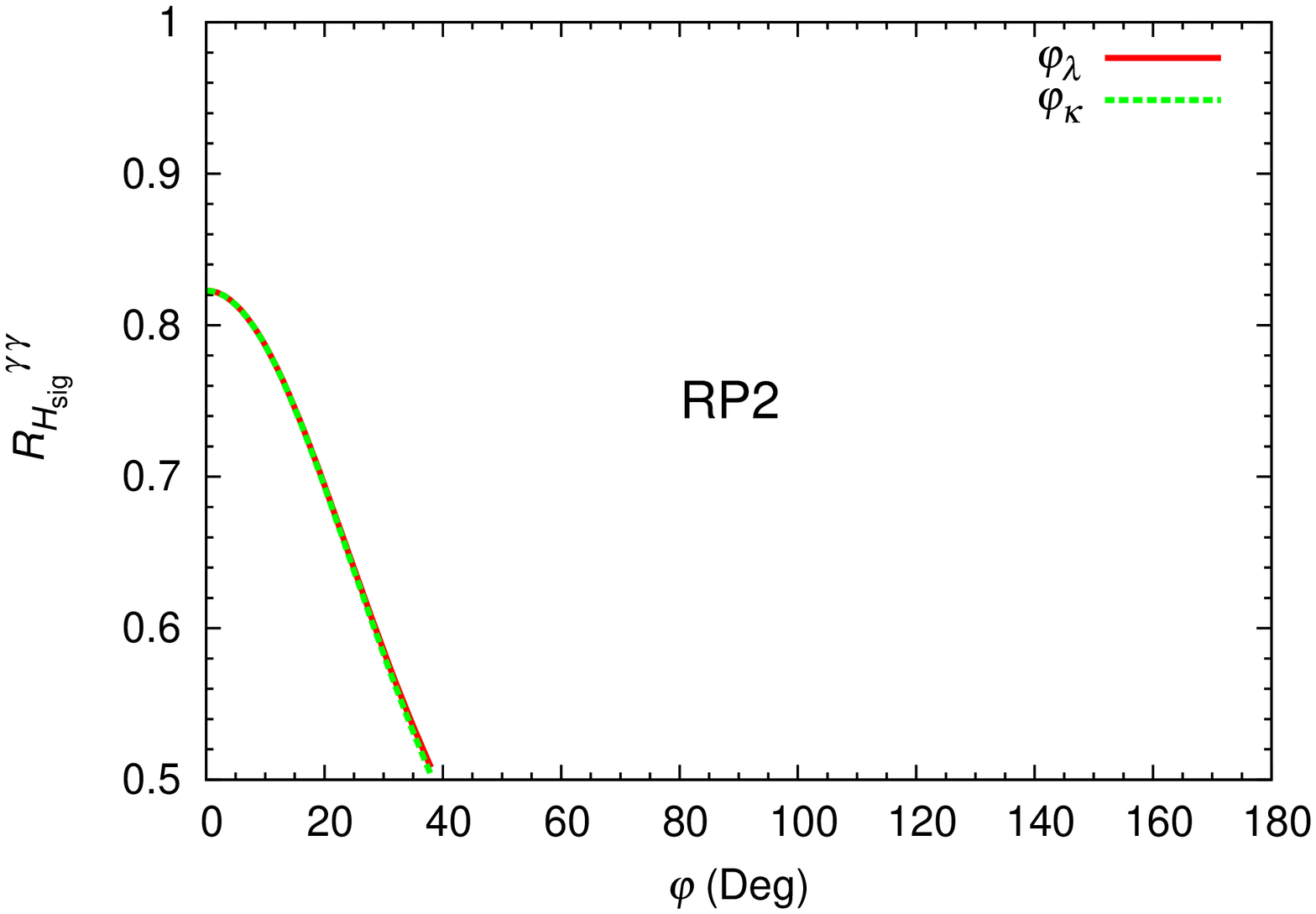}
}
\subfloat[]{%
\includegraphics*[angle=-90, scale=0.4]{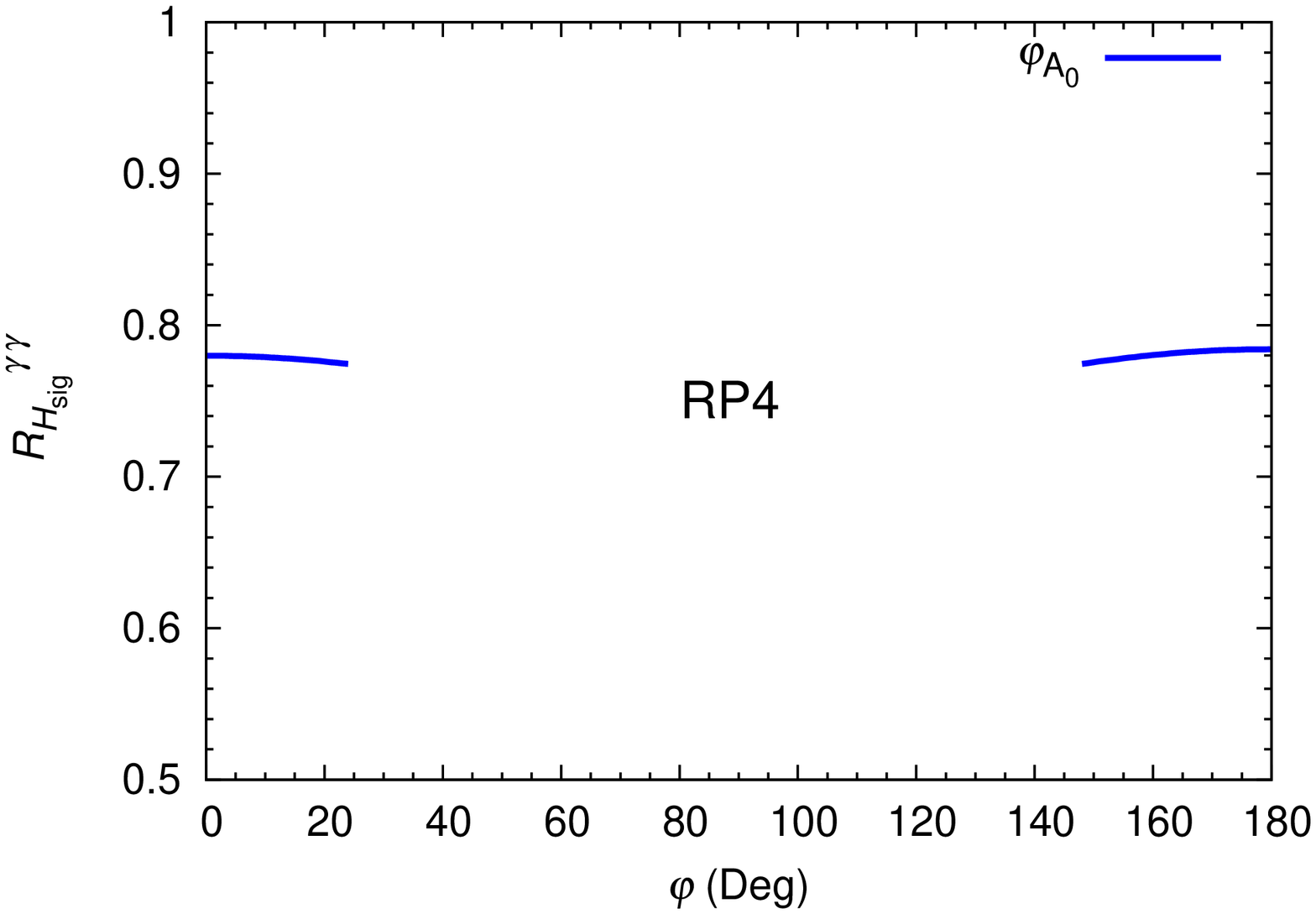}
}
\caption{Distributions of good points, $m_{H_{\rm sig}}$ and $R^{\gamma\gamma}_{H_{\rm sig}}$ for Scenario\,3.}
\label{fig:S3}
\end{figure}

Although, as noted earlier, the parameter space corresponding to this
Scenario overlaps a little with the Case\,2 of Scenario\,2, some
significant differences are noticeable. Fig.\,\ref{fig:S3}a shows a
fluctuation in the number of surviving points with varying \phlam\
and\phkap\ which is unlike any of the cases discussed so far and is
unique to this Scenario. The number of good points first rises sharply
and then falls continuously for \phkap\ between
$\sim$20$^\circ$ and $\sim$80$^\circ$. It then stays almost constant
($\sim$100) until $\phkap\simeq 120^\circ$ and then starts rising sharply
again. Note that the number of surviving points never falls to 0 for
\phkap\ while it does so for \phlam\ between 44$^\circ$ and 135$^\circ$, 
for the same reason as in Case\,2 of both Scenarios 1 and 2. Outside this gap the
lines corresponding to \phlam\ and \phkap\ overlap each other almost
completely. Although the behavior of the number of
surviving points in this Scenario is unique, the reason for it is
in fact a behavior of both \mhsig\
and \rhsiggg\ similar, but much more pronounced, to that observed in Case\,2
of Scenario\,2, as we shall explain below. Fig.\,\ref{fig:S3}b shows a
trend similar to Scenario\,2 for the number of surviving points with
increasing \phtri, except that the former dips to 0 for a small
intermediate range of the latter. The representative
points of this Scenario have the following coordinates. 
\begin{table}[h]
\begin{center}
\begin{tabular}{c|c|c|c|c}
 Point & $\lambda$ & $\kappa$ & $A_\lambda$\,(GeV) & $A_\kappa$\,(GeV)
 \\
\hline
RP1 & 0.195 & 0.09 & 1050 & -74.5\\
\hline
RP2 & 0.226 & 0.06 & 972 & -90.0 \\
\hline
RP3 & 0.226 & 0.10 & 950 & -81.1 \\
\hline
RP4 & 0.216 & 0.08 & 950 & -85.6 \\
\end{tabular}
\end{center}
\end{table}

Fig.\,\ref{fig:S3}c for RP1 shows that \mhsig\ rises very sharply with
increasing \phlam\ and \phkap\ compared to the corresponding point of
Case\,2, Scenario\,2, which is a consequence of comparatively larger
absolute values of $\lambda$ (implying larger
singlet-doublet mixing) and $\kappa$. In Fig.\,\ref{fig:S3}d for RP3 a
very typical behavior of \mhsig\ is seen with increasing \phtri. 
Fig.\,\ref{fig:S3}e for RP2 shows a behavior of \rhsiggg\ much more analogous
to that seen for RP2 of Case\,1 than of
Case\,2 of Scenario\,2 due, again, to the large absolute value of $\lambda$
involved. Note, however, the comparatively much smaller value, when CP
is conserved, of \rhsiggg, which further falls sharply with increasing amount
of CP-violation. A small value of \rhsiggg\ and minimal variation in
it with varying \phtri\ (in its range allowed by the condition on
\mhsig) is also seen for RP4 in Fig.\,\ref{fig:S3}f. Although the
maximum values of \rhsiggg\ obtainable for this Scenario still agree well with
the central value measured at the CMS, the fact that it drops sharply
with nonzero CPV phases implies that quite like the other parameters
corresponding to this Scenario the CPV phases are also much more fine-tuned
than the other two Scenarios. 

The very sharp rise in \mhsig\ and the steep
drop in \rhsiggg\ for most of the points are together responsible for
the overall behavior of the
number of surviving points noticed above. For small values of
\phlam/\phkap\, $m_{H_3}$ 
rises above the imposed lower limit on \mhsig\ for some points which
violate this limit in the CPC case. However, while $m_{H_3}$ increases \rhsiggg\
corresponding to these points falls with increasing \phlam/\phkap. 
Beyond a certain value of \phlam/\phkap\
not only do the points which previously fell within the limits on \mhsig\
and \rhsiggg\ start falling out but also the potential new points
which now have \mhsig\ above the lower limit, have too low a \rhsiggg\ 
to satisfy the limit on it. Hence, the overall number of surviving
points starts dropping. After $\phlam/\phkap=90^\circ$ the behavior of
$m_{H_3}$ is typically reversed, implying that it should start dropping
slowly again. Consequently $m_{H_3}$ for more and more points should fall
back below 127\,GeV as the values of $\phlam/\phkap$ are increased
further (although it does not happen for our RP1 here), causing the
number of surviving points to surge again. 

As with all the other cases with a $H_{dR}$-like $H_{\rm
 sig}$, \rhsigzz\ has similar values as \rhsiggg\ when the CPV phases are zero and
a similar behavior when these phases are varied. 
Finally, some details relevant to the four RPs of
this Scenario are given in Table \ref{tab:S2S3}. 

\begin{table}[t]
\begin{center}
\begin{tabular}{|l|c|c|c|}
\hline
Scenario & 2,\,Case\,1 & 2,\,Case\,2 & 3 \\
\hline
\hline
Points scanned for each $\phi_\kappa$ or $\phi_\lambda$ or
$\phi_{A_0}$ &\multicolumn{3}{|c|}{40000} \\
\hline
Points surviving for $\phi_\kappa =\phi_\lambda = \phi_{A_0}  = 0$ &
5377 & 8506  & 5944 \\
\hline 
\hline
Min. points surviving, with $\phi_\kappa$ & 0,\,14-180 & 7979,\,180
& 97,\,92-96\\
Max. points surviving, with $\phi_\kappa$ &5377,\,0
& 8506,\,0 & 8763,\,17 \\
Min. points surviving, with $\phi_\lambda$ & 0,\,14-180 & 0,\,85-95 & 0,\,44-135\\
Max. points surviving, with $\phi_\lambda$ & 5377,\,0 &
8506,\,0 & 8756,\,17 \\
\hline

$m_{h_{\rm sig}}$ for RP1 with $\phi_\kappa =\phi_\lambda = 0$ &
126.8599 & 125.9949 & 124.0 \\
\rhsiggg\ for RP2 with $\phi_\kappa =\phi_\lambda = 0$ & 1.1976 &
0.9347 & 0.8228 \\
\hline

Min. $m_{h_{\rm sig}}$ obtained for RP1, with $\phi_\kappa$ &
124.0242,\,7 & 125.9949,\,0 & 124.0,\,0 \\
Max. $m_{h_{\rm sig}}$ obtained for RP1, with $\phi_\kappa$ &
126.8599,\,0 & 126.5612,\,180 & 126.9469,\,14\\
Min. \rhsiggg\ obtained for RP2, with $\phi_\kappa$ &
0.5850,\,12 & 0.9259,\,91-98 & 0.5033,\,38\\
Max. \rhsiggg\ obtained for RP2, with $\phi_\kappa$ &1.1976,\,0 &
0.9347,\,0 & 0.8228,\,0\\
\hline

Min. $m_{h_{\rm sig}}$ obtained for RP1, with $\phi_\lambda$ &
124.0177,\,7  & 125.9949,\,0 & 124.0,\,0 \\
Max. $m_{h_{\rm sig}}$ obtained for RP1, with $\phi_\lambda$ &
126.8599,\,0  & 126.5107,\,180 & 126.9466,\,14 \\
Min. \rhsiggg\ obtained for RP2, with $\phi_\lambda$ &
0.5916,\,12 & 0.9202,\,121 & 0.5082,\,38 \\
Max. \rhsiggg\ obtained for RP2, with $\phi_\lambda$ & 1.1976,\,0 &
0.9347,\,0 & 0.8228,\,0 \\
\hline 
\hline

Min. points surviving, with $\phi_{A_0}$ &1246,\,92 & 1037,\,89 &
0,\,76-100 \\
Max. points surviving, with $\phi_{A_0}$ & 6372,\,180 & 9120,\,180 &
9241,\,180 \\
\hline
$m_{h_{\rm sig}}$ for RP3 with $\phi_{A_0}$ = 0 & 126.9997 & 126.4522
& 126.2999 \\
\rhsiggg\ for RP4 with $\phi_{A_0} = 0$ & 1.0014 & 0.7999 & 0.7799 \\
\hline
 Min. $m_{h_{\rm sig}}$ obtained for RP3, with $\phi_{A_0}$ &
 124.0516,\,124 & 124.8208,\,87 & 124.0025,\,52\\
 Max. $m_{h_{\rm sig}}$ obtained for RP3, with $\phi_{A_0}$ &
 127.0,\,159 &  126.8997,\,180 & 126.7063,\,180 \\
Min. \rhsiggg\ obtained for RP4, with $\phi_{A_0}$ & 0.7304,\,88 &
0.7731,\,138 & 0.7744,\,24 \\
 Max. \rhsiggg\ obtained for RP4, with $\phi_{A_0}$ &
1.5,\,0 & 0.8076,\,180 & 0.7842,\,180 \\
\hline 

\end{tabular}
\caption{Scan results for Scenario\,2, Cases 1 and 2, and Scenario\,3. All angles are in degrees.}
\label{tab:S2S3}
\end{center}
\end{table}


\section{Summary}
\label{sec:summa}

In summary, we have demonstrated that the CPV NMSSM offers some interesting
solutions to the LHC Higgs boson data, which differ
substantially from well-known configurations of the CPC NMSSM, thereby
augmenting the regions of parameter space which can be scrutinized at
the CERN collider. We have concentrated on the case in which only
three CPV phases, $\phi_\kappa$, $\phi_\lambda$
and $\phi_{A_0}$, enter the Higgs sector. We have
then checked the twofold impact of these
phases, always varied independently from each other, on the mass as well as signal
strength of the assumed signal Higgs boson in the $\gamma\gamma$ decay mode, 
in different model configurations.

The overall picture that emerges is that any of the three lightest
Higgs states of the CPV NMSSM can be the one discovered at the LHC. 
We have illustrated this by using five benchmark cases in the
parameter space of the model that can
easily be adopted for experimental analyses. Our analysis also proves that
the possibility of explicitly invoking CPV-phases is not ruled out by the current LHC
Higgs boson data in any of our tested plausible NMSSM Scenarios.
 Finally, a numerical tool for analyzing the Higgs sector of the CPV NMSSM has also
been produced and is available upon request. The obvious outlook of this
analysis will be to consider the possibility that companion Higgs
boson signals to the 
one extracted at the LHC may emerge in the CPV NMSSM, so as to put the LHC
collaborations in the position of confirming or disproving this SUSY
hypothesis. An investigation on these lines is now in progress. 

\section*{Acknowledgements} S. Moretti is supported in part through the NExT Institute.
S. Munir is funded in part by the Welcome
Programme of the Foundation for Polish Science. The work of P. Poulose
is partly supported by a SERC, DST (India) project, Project No. SR/S2/HEP-41/2009.

\appendix
\numberwithin{equation}{section}
\section{Mass Matrices}

Detailed expressions for the one-loop Higgs boson mass matrices can be found
in Refs. \cite{Cheung:2010ba,Graf:2012hh,Funakubo:2004ka}. Here we only reproduce the tree-level mass
matrix to show the dependence on \phlam\ and \phkap\, since
the dominant contributions from these phases arise at this level. Note
that the tree-level sfermion, neutralino and chargino mass matrices
given below are complex by definition. The one-loop effective Higgs potential
receives further contributions from \phlam\ and \phtri\ through the
squark and stau sectors and from \phlam\ and \phkap\ through the
neutralino and chargino sectors. 

\begin{itemize}

\item The neutral Higgs boson mass matrix may be written as:

\begin{eqnarray}
\label{eq:mhiggs}
\hspace{-1.0cm}
{\cal M}^2_H=
\left(
        \begin{array}{cc}
\mathcal{M}_S^2 & \mathcal{M}^2_{SP} \\
(\mathcal{M}^2_{SP})^T & \mathcal{M}_P^2  
                \end{array}
\right)\,.
\end{eqnarray}
Using the minimization conditions of the Higgs potential, one can
define some convenient parameters,
\begin{eqnarray}
\mathcal{R} &=& |\lambda| |\kappa|\, \cos(\phi^\prime_\lambda-\phi^\prime_\kappa)\,,
\hspace{1.0cm}
\mathcal{I} = |\lambda| |\kappa|\, \sin(\phi^\prime_\lambda-\phi^\prime_\kappa)\,,
\nonumber \\
R_\lambda &=& \frac{|\lambda| |A_\lambda|}{\sqrt{2}}\,
\cos(\phi^\prime_\lambda+\phi_{A_\lambda})\,,
\ \ \
R_\kappa = \frac{|\kappa| |A_\kappa|}{\sqrt{2}}\,
\cos(\phi^\prime_\kappa+\phi_{A_\kappa})\,,
\end{eqnarray}
with
\begin{equation}
\phi^\prime_\lambda \equiv \phi_\lambda+\theta+\varphi \ \ \ {\rm and} \ \ \
\phi^\prime_\kappa \equiv \phi_\kappa+3\varphi\,.
\end{equation}

In terms of these parameters, the entries of the top left $3\times 3$
CP-even block in Eq.\,(\ref{eq:mhiggs}) are given as
\begin{eqnarray}
({\cal M}_S^2)_{11}&=&\frac{g_2^2+g_1^2}{4}v_d^2
        +\left(R_\lambda+\frac{1}{2}\mathcal{R}v_S\right)\frac{v_uv_S}{v_d}, \nonumber \\
({\cal M}_S^2)_{22}&=&\frac{g_2^2+g_1^2}{4}v_u^2
        +\left(R_\lambda+\frac{1}{2}\mathcal{R}v_S\right)\frac{v_dv_S}{v_u}, \nonumber \\
({\cal M}_S^2)_{33}&=&R_\lambda\frac{v_dv_u}{v_S}+2|\kappa|^2v^2_S-R_\kappa
v_S, \nonumber \\
({\cal M}_S^2)_{12}&=&({\cal M}_S^2)_{21}
        =\left(-\frac{g_2^2+g_1^2}{4}+|\lambda|^2\right)v_dv_u
        -\bigg(R_\lambda+\frac{1}{2}\mathcal{R}v_S\bigg)v_S, \nonumber \\
({\cal M}_S^2)_{13}&=&({\cal M}_S^2)_{31}
        =-R_\lambda v_u+|\lambda|^2v_dv_S-\mathcal{R}v_uv_S,\nonumber \\
({\cal M}_S^2)_{23}&=&({\cal M}_S^2)_{32}
        =-R_\lambda v_d+|\lambda|^2v_uv_S-\mathcal{R}v_dv_S.
\end{eqnarray}

where $g_1$ and $g_2$ are the $U(1)_Y$ and $SU(2)_L$ gauge couplings,
respectively, and the bottom right $2\times 2$ CP-odd block in reads  
\begin{eqnarray}
{\cal M}^2_P=
\left(
        \begin{array}{cc}
 (R_\lambda+\frac{1}{2}\mathcal{R}v_S)\frac{v^2v_S}{v_dv_u}
& (R_\lambda -\mathcal{R}v_S)v \\
 (R_\lambda -\mathcal{R}v_S)v
 & R_\lambda\frac{v_dv_u}{v_S}+2\mathcal{R}v_dv_u-3R_\kappa v_S 
                \end{array}
\right)\,.
 \end{eqnarray}

Finally, the entries of the off-diagonal block in
Eq.\,(\ref{eq:mhiggs}), which are responsible for mixing between
CP-even and CP-odd states, are given as
\begin{eqnarray}
{\cal M}^{2}_{SP_\beta}=
\left(
        \begin{array}{cc}
0 & -\frac{3}{2}\mathcal{I}sv_u \\
0 & -\frac{3}{2}\mathcal{I}sv_d \\
\frac{1}{2}\mathcal{I}sv & -2\mathcal{I}v_uv_d 
                \end{array}
\right)\,.
\end{eqnarray}

\item The chargino mass matrix, in the $(\widetilde{W}^-,\,\widetilde{H}^-)$ basis, using the
convention $\widetilde{H}^{-}_{L(R)} = \widetilde{H}^{-}_{d(u)}$, can be written
as
\begin{eqnarray}
{\cal M}_C = \left(\begin{array}{cc}
     M_2              & \sqrt{2} m_W \cos\beta \\[2mm]
\sqrt{2} m_W \sin\beta &
\frac{|\lambda| v_S}{\sqrt{2}}\,e^{i\phlam'}
             \end{array}\right)\, ,
\end{eqnarray}
where $M_1$ and $M_2$ are the soft gaugino masses and $m_W$ is the
mass of the $W$ boson. The above matrix is diagonalized by two different unitary matrices as
$ C_R {\cal M}_C C_L^\dagger ={\sf diag}\{m_{\widetilde{\chi}^\pm_1},\,
m_{\widetilde{\chi}^\pm_2}\}$, where
$m_{\widetilde{\chi}^\pm_1} \leq m_{\widetilde{\chi}^\pm_2}$.

\item The neutralino mass matrix, in the
$(\widetilde{B},\,\widetilde{W}^0,\,\widetilde{H}^0_d,\,\widetilde{H}^0_u,\,\widetilde{S})$ basis, can be written as
\begin{eqnarray}
{\cal M}_N=\left(\begin{array}{ccccc}
  M_1       &      0          &  -m_Z \cos\beta s_W  & m_Z \sin\beta s_W  & 0\\[2mm]
 &     M_2         &   m_Z \cos\beta c_W  & -m_Z \sin\beta c_W & 0\\[2mm]
  & & 0 &
-\frac{|\lambda| v_S}{\sqrt{2}}\,e^{i\phlam'} &
-\frac{|\lambda| v s_\beta}{\sqrt{2}}\,e^{i\phlam'} \\[2mm]
 & &  &       0          &
-\frac{|\lambda| v \cos\beta}{\sqrt{2}}\,e^{i\phlam'}\\[2mm]
& & & &
\sqrt{2} |\kappa| v_S \,e^{i\phkap'}
                  \end{array}\right).\,
\end{eqnarray}
\noindent with $m_Z$ being the $Z$ boson mass, $s_W = \sin\theta_W$ and $c_W = \cos\theta_W$. This matrix is diagonalized as
$N^* {\cal M}_N N^\dagger = {\sf diag}\,
(m_{\widetilde{\chi}_1^0},\,m_{\widetilde{\chi}_2^0},\,
m_{\widetilde{\chi}_3^0},\,m_{\widetilde{\chi}_4^0},\,m_{\widetilde{\chi}_5^0})$,
where $N$ is a unitary matrix and
$m_{\widetilde{\chi}_1^0} \leq m_{\widetilde{\chi}_2^0} \leq m_{\widetilde{\chi}_3^0}
\leq m_{\widetilde{\chi}_4^0} \leq m_{\widetilde{\chi}_5^0}$.

\item For the stop, sbottom and stau matrices, in the $\left(\widetilde{q}_L,\,
\widetilde{q}_R\right)$ basis, we have

\begin{eqnarray}
\hspace*{-0.9cm}
\widetilde{\cal M}^2_t  = \left( \begin{array}{cc}
M^2_{\widetilde{Q}_3}\, +\, m^2_t\, +\, \cos 2\beta m^2_Z\, (
\frac{1}{2} - \frac{2}{3} s_W^2 ) &
h_t^* v_u (|A_t| e^{-i(\theta+\phi_{A_t})} -
\frac{|\lambda| v_S}{\sqrt{2}}e^{i\phlam'} \cot\beta )/\sqrt{2}\\
h_t v_u (|A_t| e^{i(\theta+\phi_{A_t})} -
\frac{|\lambda| v_S}{\sqrt{2}}e^{-i\phlam'} \cot\beta )/\sqrt{2}
& \hspace{-0.2cm}
M^2_{\widetilde{U}_3}\, +\, m^2_t\, +\, \cos 2\beta m^2_Z\, Q_t s^2_W
\end{array}\right)\,, \nonumber
\end{eqnarray}

\begin{eqnarray}
\hspace*{-0.9cm}
\widetilde{\cal M}^2_b  = \left( \begin{array}{cc}
M^2_{\widetilde{Q}_3}\, +\, m^2_b\, +\, \cos 2\beta m^2_Z\, (
-\frac{1}{2} + \frac{1}{3} s_W^2 ) &
h_b^* v_d (|A_b|e^{-i\phi_{A_b}} -
\frac{|\lambda| v_S}{\sqrt{2}}e^{i\phlam'} \tan\beta )/\sqrt{2}\\
h_b v_d (|A_b|e^{i\phi_{A_b}} -
\frac{|\lambda| v_S}{\sqrt{2}}e^{-i\phlam'} \tan\beta )/\sqrt{2}
& \hspace{-0.2cm}
M^2_{\widetilde{D}_3}\, +\, m^2_b\, +\, \cos 2\beta m^2_Z\, Q_b s^2_W
\end{array}\right)\,,  \nonumber
\end{eqnarray}

\begin{eqnarray}
\hspace*{-0.9cm}
\widetilde{\cal M}^2_\tau  = \left( \begin{array}{cc}
M^2_{\widetilde{L}_3}\, +\, m^2_\tau\, +\, \cos 2\beta m^2_Z\,
( s_W^2-1/2 ) &
h_\tau^* v_d (|A_\tau|e^{-i\phi_{A_\tau}} -
\frac{|\lambda| v_S}{\sqrt{2}}e^{i\phlam'} \tan\beta )/\sqrt{2}\\
h_\tau v_d (|A_\tau|e^{i\phi_{A_\tau}} -
\frac{|\lambda| v_S}{\sqrt{2}}e^{-i\phlam'} \tan\beta )/\sqrt{2}
& \hspace{-0.2cm}
M^2_{\widetilde{E}_3}\, +\, m^2_\tau\, -\, \cos 2\beta m^2_Z\, s^2_W
\end{array}\right)\,,
\end{eqnarray}

\noindent where $m_t$, $m_b$ and $m_{\tau}$ are the masses of $t,\,b$ quarks 
and $\tau$ lepton, respectively, and $y_t$,
$y_b$ and $y_{\tau}$ are the corresponding Yukawa couplings. $Q_t$ and $Q_b$ are the respective
electric charges of the $t$ and $b$ quarks. The mass
eigenstates of top and bottom squarks and stau are obtained by diagonalizing the above mass matrices 
as $U^{\tilde{f}\dagger} \, \widetilde{\cal M}^2_f \,
U^{\tilde{f}} ={\sf
  diag}(m_{\tilde{f}_1}^2,m_{\tilde{f}_2}^2)\,$, such that
$m_{\tilde{f}_1}^2 \leq m_{\tilde{f}_2}^2$, for $f=t,\,b,\,\tau$.

\end{itemize}

\end{document}